\documentclass[12pt]{article}
\usepackage{amsthm,amsfonts
,
}
\title{Recent developments in
quantum mechanics with magnetic fields}
\author{L\'aszl\'o Erd\H os\thanks{Partially
supported by  EU-IHP Network ``Analysis
and Quantum'' HPRN-CT-2002-0027 and by Harvard University}}

\date{October 28, 2005}

\newcommand{\rd}{{\rm d}}
\newcommand{\be}{\begin{equation}}
\newcommand{\ee}{\end{equation}}
\newcommand{\bey}{\begin{eqnarray}}
\newcommand{\eey}{\end{eqnarray}}
\newcommand{\beys}{\begin{eqnarray*}}
\newcommand{\eeys}{\end{eqnarray*}}

\newcommand{\bsigma}{\mbox{\boldmath $\sigma$}}

\newcommand{\balpha}{\mbox{\boldmath $\alpha$}}

\newcommand{\bR}{{\bf R}}
\newcommand{\bC}{{\bf C}}

\newcommand{\e}{\varepsilon}

\newcommand{\tr}{{\rm tr}}
\newcommand{\sfrac}[2]{{\textstyle \frac{#1}{#2}}}

\newcommand{\cD}{{\cal D}}

\newcommand{\cK}{{\cal K}}

\begin{document}
\maketitle

\begin{abstract}
We present a review on the
recent developments concerning rigorous
mathematical results on Schr\"odinger operators with magnetic fields.
This paper is dedicated to  the sixtieth birthday of Barry Simon.
\end{abstract}


\medskip\noindent

\medskip\noindent
{\it Running title:} Recent developments on magnetic fields


\section{Introduction}\label{sec:intro}

The mathematical formulation of 
quantum mechanics, given by Schr\"odinger, Pauli and Dirac,
has posed 
an enormous challenge: can mathematics, with
its own tools and standards, rigorously justify or even
predict  physical phenomena of the quantum world?
Similarly to the development of the differential and integral
calculus, strongly motivated by Newton's classical mechanics,
new mathematical tools have been created (most notably
by von Neumann, Weyl, Wigner and later by Kato).
Functional analysis, representation theory  and 
partial differential equations would have been much poorer
mathematical disciplines without quantum mechanics.

Electromagnetic fields play a central role in quantum physics;
their rigorous inclusion in the theory is certainly one of the
key goals of mathematical physics.
Quantum electrodynamics (QED) postulates that electric and magnetic
fields are to be described within a unified relativistic theory.
Although the framework for QED
has been  clear since the 30's, 
the mathematical difficulties even to formulate the theory
rigorously have not yet been resolved.
In the low energy regime, however, massive quantum particles 
can be described non-relativistically. 
Electric and magnetic fields, with a good approximation,
can be considered decoupled.
Since typical magnetic fields in laboratory
are relatively weak, as a first approximation one can completely
neglect magnetic fields and concentrate only on
quantum point particles interacting via electric potentials.

The rigorous mathematical theory of Schr\"odinger operators
has therefore started with studying the operator
$H= \frac{1}{2m} p^2 + V(x)$ on $L^2(\bR^d)$ and its multi-particle analogues.
Here $x\in \bR^d$ is the location of the particle in
the $d$-dimensional configuration space, $p=-i\nabla_x$
is the momentum operator and $m$ is the mass, that can be
set $m=\frac{1}{2}$ with convenient units.
The Laplace operator describes the kinetic energy of the particle
and the real-valued function $V(x)$ is the electric potential.
Although both the kinetic and potential energy operators
are very simple to understand separately,
their sum exhibits a rich variety of complex phenomena
which differ from their classical counterparts
in many aspects. The mathematical theory of this operator is 
the most developed and most extensive in mathematical physics:
 the best recent review is by  Simon  \cite{S-00}.

As a next approximation, classical magnetic fields are included
in the theory, but spins are neglected.
  The kinetic energy operator is modified
from $p^2$ to $(p+A)^2$ by the minimal substitution rule:
$p\mapsto p-eA$ and we set the charge to be $e=-1$.
 Here $A: \bR^d\to \bR^d$
 is the magnetic vector potential that generates the magnetic
field $B$ according to classical electrodynamics.
 In $d=2$ or $d=3$
 dimensions $B= \nabla \times A$ is a scalar or a vector field,
respectively. In $d=1$ dimension the vector potential
can be removed by a unitary gauge transformation, $e^{i\varphi} (p+A)^2
e^{-i\varphi} = p^2$, $\varphi=\int A$, therefore  magnetic phenomena
in $\bR^1$  are absent (they are present in the case of $S^1$).

We will call the operator $(p+A)^2 +V$ 
the {\it magnetic Schr\"odinger operator.} In general, even the
kinetic energy part contains  noncommuting
operators, $[(p+A)_k, (p+A)_\ell]\neq 0$, and the theory of 
$(p+A)^2$ itself is more complicated than that of $p^2 +V$.
The simplest case of constant magnetic field, $B=const$,
is  explicitly solvable. The resulting   Landau-spectrum consists,
 in two dimensions, of infinitely 
degenerate eigenvalues at  energies $(2n+1)|B|$, $n=0,1,\ldots$.
Notice that the magnetic spectrum is characteristically different
from that of the free Laplacian. The eigenfunctions
are localized on a scale $|B|^{-1/2}$; this corresponds
to the cyclotronic radius in classical mechanics
(Landau orbits).

The interaction of the spin with a magnetic field is proportional
to the field strength. In the low energy regime this effect
is comparable with the energy shift due to 
inclusion of $A$ into $p^2$. 
Since electrons are spin-$\sfrac{1}{2}$ particles, the spin, in
principle, should not be neglected whenever magnetic fields
are considered.  Nevertheless, magnetic Schr\"odinger operators
constitute an important intermediate step to understand
magnetic phenomena.

The state space of a spin-$\sfrac{1}{2}$ particle
is $L^2(\bR^d, \bC^2)$ (in $d=2,3$) and the
momentum operator is the {\it Dirac operator}, $\cD_A:=\bsigma\cdot (p+A)$,
where $\bsigma = (\sigma_1, \sigma_2, \sigma_3)$ is the
vector of the Pauli matrices. The kinetic energy
is given by the {\it Pauli operator},
\be
    H_{P} := \cD_A^2 = [\bsigma\cdot (p+A)]^2 = (p+A)^2 + \bsigma \cdot B,
\label{lich}
\ee
and external potential may be  added as before.
The last identity is a special case of the Lichnerowicz formula
known from spinor-geometry.

In most of this review we restrict ourselves to these
operators and their multi-particle generalizations. 
However, we briefly mention that 
in dimensions higher than 3  or on
configuration spaces with a non-flat Riemannian metric,
the vector potential is canonically defined
as a one-form, $\alpha$, and the magnetic 2-form,
$\beta=\rd \alpha$, is its exterior derivative.
In the conceptually most general setup for the spinless
case, the Hilbert
space of states consists of the $L^2$-sections of a $U(1)$-bundle
over  an orientable Riemannian manifold, $M$, representing the configuration
space, and the momentum operator is the covariant derivative, $\nabla$,
on this bundle. In this formulation, the vector potential does
not appear directly but the magnetic field is ($i$-times)
the curvature 2-form of $\nabla$.  
Proper description of the spin involves covariant derivatives on 
sections  of an Spin${}^c$-bundle with Pauli matrices replaced
by Clifford multiplication \cite{ES-III}.

\medskip

In relativistic theories, electron-positron pair-creations 
cannot be neglected and one studies the full {\it relativistic
Dirac operator}, $\balpha\cdot (p+A) + \beta m$, where
$(\balpha, \beta)$ is the vector of the Dirac  matrices
and $m$ is the mass. Due to the lack of semiboundedness of the Dirac
operator, its definition, even without a
magnetic field, is a complex issue that is not
yet satisfactorily resolved in the many-body situation
 (``filling the Dirac sea''). 
We will not pursue this direction here since
the current research focuses more on the non-magnetic 
aspects of the Dirac operator.

A consistent quantum theory  requires to quantize
the electromagnetic field as well. Ideally, this should be
done within the framework of the Dirac operator (relativistic QED) but 
this problem is beyond the reach of the current techniques.
A  more tractable model is the nonrelativistic QED, where quantized
electromagnetic field is introduced in the Pauli operator,
i.e. pair-creations are neglected.

\bigskip

This overview gives an admittedly biased
summary of a few recent key results involving 
magnetic Hamiltonians. Many people have contributed
to these questions and a selection was unavoidable;
the author apologizes to everyone whose work have been 
left out. The choice reflects the author's taste
and the pressure of the editors to keep 
the page limit. 

In Section \ref{sec:def}
we present results related to the proper definitions of these operators.
In Section \ref{sec:1} we discuss 
one-particle spectral theory, including
 Lieb-Thirring type bounds and semiclassical 
methods. In Section \ref{sec:mult}
we consider multi-particle problems, including stability
of matter, large atoms  and scattering. Finally, Section \ref{sec:random}
is devoted to random Schr\"odinger operators with magnetic fields.

Barry Simon was undoubtedly one of the initiators
and most important contributors of the endeavor
to put  Schr\"odinger operators 
on a solid mathematical ground. His work was especially 
pionieering in the theory of magnetic fields. Among
many of his achievements in this area, here I would just 
mention those two that had the biggest impact 
on my own work. Barry was the first who
systematically exploited path integral methods
for magnetic fields upon an initial suggestion
of Nelson (see e.g. \cite{S-79}). Secondly,
his seminal papers with Avron and Herbst \cite{AHS}
have become the   classical reference ``handbook''
about magnetic fields. This overview is dedicated to
his 60-th birthday.

\section{Basic  qualitative properties}\label{sec:def}

\subsection{Definitions}

Along the development of the rigorous
theory of Schr\"odinger operators without magnetic fields,
it was apparently  Kato 
who first initiated  the natural program to extend this theory
to the most general magnetic fields. The unique self-adjoint
extension of the operator $(p+A)^2 + V$
without any growth condition on $A$ was
shown in 1962 by Ikeda and Kato \cite{IK}.
This result indicated that magnetic operators
should not simply be  viewed as second order differential
operators with variable coefficients.
For most mathematical purposes it is misleading
to look at $(p+A)^2$ as $p^2 + A\cdot p + p\cdot A + A^2$.
The $A$-field plays a special role in magnetic problems:
it balances the derivative  of the phase of the 
wave function. This effect is inherently present in the form
$(p+A)^2$.

Kato has proved his celebrated
distributional inequality, $\Delta |\psi| \ge \mbox{Re}\;
[\mbox{sgn}\psi \; \Delta\psi]$, for any $\psi\in L^1_{loc}$,
$\Delta \psi \in L^1_{loc}$ in 1973 \cite{K-73}. 
 Simon  has realized  its connection
to the semigroup inequality, 
$|e^{t\Delta}\psi|\leq e^{t\Delta}|\psi|$ in 1977
\cite{S-77}. A more general
abstract setup was considered in \cite{S-79(b)}, and
independently in \cite{HSU}, leading to the
magnetic  versions of these
inequalities.

For regular vector potential,
a simple proof of the semigroup
diamagnetic inequality,
\be
|e^{-t(p+A)^2}\psi|\leq e^{t\Delta}|\psi|\;,
\label{diam}
\ee
 via the Feynman-Kac formula was given in
Simon's paper \cite{S-77} quoting an  argument  of Nelson
in a private communication. It was apparently Nelson
who pointed out the probabilistic approach to Simon
(see the history in Simon's book \cite{S-79}), but the
real power behind the rigorous path integral
method for magnetic fields was realized 
in a series of work of Simon and collaborators
\cite{AHS} (see also \cite{CSS}).

More singular vector potentials  were  considered with analytic
methods in \cite{K}.
Finally, in his seminal paper \cite{S-79(a)}, Simon has given a simple
proof of the diamagnetic inequality (\ref{diam})
where the operator $H=(p+A)^2$ was defined under the most general
conditions, namely for $A\in L^2_{loc}$
and for $H$ defined as the operator associated with 
the maximal quadratic form. The domain of the maximal
form contains all $\psi\in L^2$ with $(p+A)\psi \in L^2$
in distributional sense. Using (\ref{diam}) and semigroup smoothing,
Simon showed that the $C_0^\infty$ is a form core for $H$.
A non-negative potential $V\in L_{loc}^1$ can be added to $H$ without any 
difficulty.

The optimal conditions for $C_0^\infty$ being the operator
core for $H$ are $A\in L^4_{loc}$ and  $\mbox{div} A\in L^2_{loc}$. This
was conjectured by Simon and proved by
Leinfelder and Simader, \cite{LS-81}. Leinfelder \cite{Lein}
has also showed the unitarity equivalence
under any gauge transformation, $A\to A+\nabla \varphi$, that
stays within the above classes.
Again, a non-negative potential $V\in L_{loc}^2$ 
can be added to $H$ without any difficulty 
and the Leinfelder-Simader theorem extends to a certain class
of negative potentials as well ($V=V_1+V_2\leq 0$, $V_1, V_2\in L^2_{loc}$,
$V_1(x)\ge -(const.)|x|^2$, $V_2$ bounded relative to $-\Delta$
with a bound smaller than one).

The most general conditions on 
potentials with non-trivial negative part, $V_-\neq 0$,
are  hard to use directly. The typical argument uses the KLMN theorem
(Theorem X.17 \cite{RSi}) 
that defines self-adjoint operators by adding a relative
form bounded perturbation (with bound less than 1)
to a semi-bounded closable quadratic form. 
The boundedness of $V_-$ relative to $(p+A)^2 +V_+$
 is, however,  hard to check. With the help
of the diamagnetic inequality, the boundedness of $V_-$
relative to $p^2 +V_+$ is sufficient. We recall that
$V_-$ being in the Kato class, $V_-\in \cK$, implies
even infinitesimal boundedness.  
Using  Stratonovich stochastic integrals, the Feynman-Kac formula
can be extended to $A\in L^2_{loc}$ vector potentials
if $V_-$ is relative bounded by $p^2 +V_+$ (\cite{Hun}).
If one prefers to use  Ito stochastic integrals, then,
additionally, $\mbox{div} A\in L_{loc}^2$ is also necessary
for the Feynman-Kac formula. 
The definition of the magnetic operator with Neumann
boundary conditions was carefully worked out recently in \cite{HLMW}
and the proof 
of the diamagnetic inequality and Kato's inequality
were extended to this case using the method of Simon \cite{S-79(a)}.
The most general diamagnetic result for the Neumann case is obtained 
in \cite{HS}  that uses no regularity assumptions on the domain and on $V_+$.

Similarly to the non-magnetic case worked out
in the fundamental paper by Simon \cite{S-82},
with the help of the (magnetic) Feynman-Kac formula one
can prove smoothing and continuity properties
of the semigroup and its kernel. This work has been carried
out in \cite{BHL} with great care and with many fine details.
To summarize the results, one assumes that 
the vector potential $A$  belongs to the so-called
(local) magnetic Kato class, i.e. $A^2, \mbox{div} A \in \cK_{loc}$,
and the potential is Kato decomposable ($V_+\in \cK_{loc}$, $V_-\in \cK$).
Then the $L^p$-semigroup is continuous in time and if $A$ and $V$
are approximated locally in the Kato-norm, then the approximating
semigroups converge. Moreover, the Feynman-Kac formula defines
a continuous representation of the semigroup kernel.

\bigskip

The definition of the Pauli operator can be directly
reduced to that of the magnetic Schr\"odinger operator
using (\ref{lich}) and treating $\bsigma\cdot B$ as
a (matrix-valued) potential term. However,
the supersymmetric structure of the Pauli operator
(at least in even dimensions) allows one to define the Pauli
operator directly and for more general magnetic fields.
On topologically trivial
domains only the magnetic field has physical relevance.
The weakest necessary condition on $\bsigma\cdot B$, if considered
as a potential, is $B\in L^1_{loc}$. However, not every $L^1$
field can be generated by an $L^2_{loc}$ vector potential, hence
$(p+A)^2$ might not be defined even as a quadratic form.
Therefore it is desirable to define the Pauli operator
directly, by circumventing the vector potential.
This idea has been worked out in $d=2$ dimensions in \cite{EV},
where  $A$ was replaced by a scalar potential, $h$, satisfying
$\Delta h=B$, and the Pauli quadratic form was given by
\be
    q(\psi, \psi):= 4\int | \partial_{\bar z}(e^{-h}\psi_+)|^2 e^{2h}
 + 4\int | \partial_{z}(e^{h}\psi_-)|^2 e^{-2h}, \qquad \psi=\pmatrix{\psi_+
\cr\psi_-} .
\label{qh}
\ee
This definition is applicable for any measure-valued
magnetic field and it is consistent with the 
standard one for fields that can be generated
by $L^2_{loc}$ vector potential. However, for singular
fields the form core is not $C_0^\infty$ any more.
Strangely enough, similar construction does not seem to apply
for the magnetic Schr\"odinger operator and the higher dimensional
generalizations are also open.

\subsection{Compact resolvent, essential spectrum, absolute continuous spectrum}

A basic qualitative fact about  magnetic fields is 
that their inclusion into the free spinless Laplacian,
very roughly saying, increases the bottom of the local spectrum
by $|B(x)|$.
This intuitive statement makes sense only
if the spectrum of the localized operator can be defined
and if $B(x)$ is sufficiently regular. The key
mathematical reason is the Lichnerowicz identity (\ref{lich})
that shows that $(p+A)^2$ on spinors is a nonnegative operator  plus
$-\bsigma\cdot B$. Viewing this identity restricted
to spinors with a spin direction opposite to the field,
one obtains a useful lower bound on the magnetic Schr\"odinger
operator. One can also see this by using 
the fact that  the components of $(p+A)^2$ do not commute:
the commutator is the magnetic field (up to a factor $i$).
 This trivial but crucial observation
is the core of many results throughout the next sections.
We emphasize that this effect holds only for the spinless
magnetic Schr\"odinger operator and not for the Pauli
operator.

This idea has been elaborated by several authors to investigate
the location of the essential spectrum and resolvent compactness
of the magnetic operators. For the Schr\"odinger operator
it has been shown that for sufficiently regular
fields the condition that the strength of the magnetic field
goes to infinity is equivalent to the compactness of the resolvent
(see \cite{HM-88} and references to previous results therein, in
particular \cite{AHS}).
The regularity assumptions were weakened in \cite{Sh-96} 
by using functions belonging to
the so-called reverse H\"older class.
For the Pauli operator without external potential it is conjectured that its
resolvent is never compact. This has been shown for
sufficiently ``well-behaving'' magnetic fields in \cite{HNW}.
Under stronger conditions about the
magnetic field at infinity,
the essential spectrum of the Pauli
operator was also identified in \cite{HNW}.
 In a recent work of Last and Simon \cite{LaS}
a different characterization of the essential
spectrum was given in terms of the union of the spectra
of certain limit operators at infinity.

\bigskip

The  localized eigenfunctions of the Landau spectrum in
case of a constant magnetic field
 indicate that in $d=2$ dimensions the magnetic field has
a strong localization effect, while in $d=3$ dimensions
the free motion parallel with the field guarantees
absolutely continuous spectrum. It is somewhat
surprising that by a small change of the constant magnetic
field, the spectrum can become purely  absolutely continuous
even in $d=2$. This was first observed and proved by Iwatsuka \cite{Iwa}
for magnetic fields that are translation invariant in
one direction and tend to two different values at 
plus and minus infinity in the other direction.
The classical analogue of this model is actually a 
very simple geometric picture. Since the cyclotronic radius depends
on the field strength, the closed Landau orbits become
spirals whose average velocity are nonzero and perpendicular
to the gradient of the field. 

Similar phenomenon can
be created by an external potential in constant magnetic field
or by Dirichlet boundary conditions along an edge of
the sample that extends to infinity. 
Under suitable conditions the states can be classified
as edge states and bulk states. The edge states are localized
along the boundary and they give rise to pure absolutely
continuous spectrum inside the Landau gaps. They carry
nonvanishing chiral edge currents. This picture persists
 even under perturbations with a small (possibly random)
potential \cite{FGW}. The edge states exhibit a 
level repulsion that is even stronger than that
of the Gaussian ensembles expected for
the usual Anderson model in the extended states regime
\cite{Ma}.

\subsection{Zero modes and multiplicity}\label{sec:zm}

The supersymmetric structure of the Pauli operator is responsible
for the spectrally rigid and
typically large kernel of $H_P$ in $d=2$. The
 Aharonov-Casher theorem \cite{AC, CFKS, Mi}
states that $\mbox{dim Ker} H_P$
is given (essentially) by the total flux, divided by $2\pi$, 
$\frac{1}{2\pi}\int B$. 
As a special case of the Index Theorem, for smooth data and 
on a compact manifold,  it basically relies
on algebraic identities.
 Still, in its most general form on $\bR^2$ it was only recently
proved in \cite{EV} (for finite total flux)
and \cite{RS} (for non-negative field) using (\ref{qh}).
For arbitrary field it is false \cite{EV}. 
In the strong field limit, under some regularity assumptions,
the local density of Aharonov-Casher zero modes converge
to $\frac{B(x)}{2\pi}$ \cite{E-93}.

The elements of the kernel the Pauli operator (the so-called
zero modes) in $d=2$ are 
conceptually much easier to understand than in $d=3$.
Naive extensions of the two dimensional constructions to three dimensions
fail, and they even seem to indicate that there are no zero modes
in $d=3$. A fundamental observation of Loss and Yau \cite{LY}
is that the equation
$$
   \cD_A\psi =0, \qquad A, B, \psi \in L^2
$$
does have a solution on $\bR^3$, albeit quite complicated.
 This seemingly innocent fact
implies, among others, that nonrelativistic matter
with a magnetic field cannot be stable unless the fine
structure constant is sufficiently small \cite{FLL}.

The explicit construction of \cite{LY} does not shed much
light on the conceptual origin of the zero modes.
It turns out that two dimensional Aharonov-Casher zero modes
on $S^2$ can be lifted to $\bR^3$ using the Hopf map
and spinor geometry \cite{ES-III} (see also  \cite{AMN, El-1}
for other examples). In particular, magnetic fields
with arbitrary number of zero modes can be constructed.
Although many zero modes are obtained
 in this conceptual way, still not all explicit
zero modes of \cite{LY} are covered. On the other side,
it is known that magnetic fields with zero modes
form a slim set in the space of all magnetic fields
\cite{BE}, \cite{El-2}. It is an interesting open question 
to connect the existence, or even the number of the
zero modes with the geometry of the magnetic field. 
Currently we have even not a conjecture for a
general characterization of magnetic fields with zero mode.

\medskip

For the spinless magnetic Schr\"odinger operator no
supersymmetric structure is available to analyze 
the ground states and even to compute the bottom of the spectrum
is complicated, apart from the strong field
regime (Section \ref{sec:ind}).  Since the Perron-Frobenius
theorem does not apply to the magnetic
Laplacian, the ground state can be degenerate, 
although for generic field it is simple.
Still, the strength of magnetic field 
restricts the possible multiplicity.
Based upon similar observations by Colin de Verdi\`ere 
on graphs, it was conjectured in \cite{CdVT}
that on a two dimensional manifold $M$,
 the total curvature of the line
bundle, i.e. the total flux, $\int_M |B|$, 
gives an upper bound on the multiplicity
of the magnetic ground state. This was proved 
in \cite{E-02} modulo constants depending on the
geometry of the base manifold. The same bound
with constants depending only on the genus of $M$
is still an intriguing open question.
The proof in \cite{E-02}
relies on an upper bound on the ground state energy
in terms of the total flux and this intermediate
result necessarily depends on the geometry of $M$.
The construction of an appropriate trial state
uses the scalar potential $h$ (with $\Delta h=B$) 
instead of the vector potential in order to 
control the energy solely by the $L^1$ norm of $B$.

\subsection{Magnetic operators on the lattice}\label{sec:latt}

The magnetic Schr\"odinger (and Pauli) operator can also be defined 
on the lattice. The magnetic field is defined
on the plaquets, while the magnetic vector potential, $A_b$,
is a function on the bonds. The magnetic translation operator
along the bond $b$
amounts to a multiplication by a complex phase $e^{iA_b}$ in addition
to the usual  hopping. The field on each plaquet is
the oriented sum of the vector potentials along the boundary.
The field and the vector potential are defined only modulo $2\pi$.

Although this definition is very natural, the spectral
properties of the lattice magnetic Schr\"odinger operator
differ vastly from the continuous version.  Even for a
constant magnetic field $B$ on a regular two dimensional
square lattice (Harper operator),
the spectrum exhibits a complex fractal behavior (``Hofstadter
butterfly'')
 sensitively depending
on the diophantine properties of $\alpha=B/2\pi$.
 With a simple
transformation, this operator can be reduced to the almost Mathieu
 operator; a simple prototype
of a one-dimensional discrete Schr\"odinger operator
with an almost periodic potential:
\be
     H_{\alpha, \lambda}= \lambda \cos (2\pi \alpha D_x) + \cos x\; .
\label{harper}
\ee
The continuous Schr\"odinger operator with a constant
magnetic field and periodic potential leads to a similar equation.

 The Cantor-like spectrum of $H_{\alpha, \lambda}$
was first proven in \cite{BeS} for a dense set of parameter values
and later in a series of papers Helffer and Sj\"ostrand
performed a detailed quantitative semiclassical analysis \cite{HS-1} 
initiated by Wilkinson \cite{Wil} to identify a large set of
parameter values $\alpha$ with Cantor spectrum if $\lambda=1$. 
With quite different techniques, Last obtained a similar result
and he  also computed the Lebesgue measure of the spectrum \cite{Last}
for all $\lambda$. Finally, the Cantor spectrum has recently been proven
 by Puig \cite{Pu} ($\lambda\neq 0$,
$\alpha$ is Diophantine) and by Avila and Jitomirskaya \cite{AJ}
for all conjectured values of the parameters: $\lambda\neq 0$,
$\alpha$ irrational (``Ten martini problem'', as it was named
and popularized by Barry Simon).

\subsection{Dia- and paramagnetism}

Diamagnetism plays a crucial role in the analysis of
the magnetic Schr\"odinger operators since it gives
an easy apriori comparison of magnetic and non-magnetic operators,
like (\ref{diam}).
However, the  apparent strength of the basic diamagnetic
inequality is somewhat misleading
when it comes to quantitative results. 

On one hand, it completely neglects magnetic effects;
operators with two different but nonzero
magnetic fields are not comparable with this method.
In particular, diamagnetism in a strong sense, i.e. monotonicity of
the energy in the magnetic field strength,
does not hold in general because of the de Haas-van Alphen
oscillation effect (see \cite{HS-2} for a rigorous
proof in the weak field regime with a periodic external potential).

On the other hand, diamagnetism is applicable only for
the exponential statistics, $\tr \; 
e^{-\beta H} = \sum_j e^{-\beta \lambda_j}$,
of the eigenvalues, $\lambda_j$, in particular for  
 the ground state ($\beta\to \infty$).
Going beyond these constraints is notoriously difficult
and there are only a few results and many open questions.

Loss and Thaller proved \cite{LT-97} 
that the heat kernel of a two dimensional Schr\"odinger
operator $H=(p+A)^2$ with 
an arbitrary magnetic field $B(x)$ can be estimated by
$$
     \Big|  e^{-tH}(x,y)\Big|
\leq \frac{B}{4\pi \sinh Bt} e^{-\frac{(x-y)^2}{4t}}
$$
if $B(x)\ge B (\ge 0)$. The right hand side 
is smaller than the free heat kernel and its exponential
behaviour $\approx e^{-Bt}$ correctly reflects a
spectral shift at the ground state energy by 
at  least $B$. However,
 it does not retain the full Gaussian offdiagonal decay
of the magnetic heat kernel with a constant field.
With the help
of this inequality, sharp $L^p-L^q$ bounds were shown in \cite{LT-97}.
 The proof heavily
uses the Gaussian character of the heat kernel of
the constant field operator.  Several counterexamples
\cite{E-97} show that this result is basically the best
one could hope for: there is no strict diamagnetic
comparison between two non-homogeneous magnetic fields
or even between two homogeneous magnetic fields with a potential.
The Gaussian offdiagonal decay cannot be fully recovered.
Such type of decay apparently requires real analyticity of
the magnetic field and potential in the angular direction
\cite{E-96} \cite{Na1} \cite{Sor}.

In the large field limit, the diamagnetic
effect is so strong that the magnetic Schr\"odinger
operator converges (in resolvent sense) to the free
Laplacian with Dirichlet boundary conditions
on the regime where the magnetic field vanishes
\cite{HH}. In other words, strong magnetic fields
act like Dirichlet walls, confining the electron
motion to regimes where the field is zero.

In contrast to the diamagnetism of the (spinless) magnetic
Schr\"odinger operator, the Pauli operator tends to be
paramagnetic. This issue was apparently raised first
in \cite{HSS} and E. Lieb proved (Appendix of \cite{AHS})
that the ground state energy of the Pauli operator with potential cannot
increase as a constant magnetic field is turned on. 
However, paramagnetism fails for non-homogeneous fields
\cite{AS} \cite{H-88} \cite{E-97}.

\medskip

For many-fermion systems, one studies the sum 
of the low lying eigenvalues 
of the one-body operator. This statistics is more singular,
it is beyond the exponential statistics offered by
the heat kernel and surprising phenomena occur.
The magnetic Schr\"odinger operator on a squarel
lattice turns out to be paramagnetic at half-filling.
It is the maximal flux ($\pi$ on each plaquet)
 that minimizes the sum of the first $\Lambda/2$ magnetic
eigenvalues on a torus of volume $\Lambda$.  
The result
actually holds on any bipartite graph that has a periodicity
at least in one direction.
After some special cases presented in \cite{L-92} and
proved in \cite{LL-93}, the general
result was proven by Lieb \cite{L-94}.
 The proof uses reflection positivity  and seemingly it  cannot
be extended to other filling factors or to graphs without
periodicity,  leaving the general case as an intriguing open question. 

Diamagnetism for sums of the Schr\"odinger eigenvalues
fails in the continuum as well. For a compact domain in $\bR^d$
and for a constant magnetic field $B$, let $\lambda_j(B)$ be
the $j$-th magnetic eigenvalue. The sum of
the first $N$ eigenvalues, $\sum_{j=1}^N \lambda_j(B)$,
may decrease by turning on a nonzero magnetic field,
but it can never drop below the semiclassical bound
\cite{ELV}. The proof heavily relies on
the homogeneity of the magnetic field. For this case
 a stronger diamagnetic inequality was proven:
$$
   \tr [ \chi f( \,(p+A)^2 \, ) ]\leq  
\tr [ \chi f( \, p^2 \, ) ] 
$$
for an arbitrary nonnegative, convex function $f$ decaying to zero at
infinity. 
Here $\chi$ is the characteristic function of an open set,
the operators are defined in the whole $\bR^n$.
This stronger diamagnetic inequality fails
for non-homogeneous fields \cite{ELV} but still the semiclassical bound
for the eigenvalue sum is conjectured to hold.

\subsection{One-body scattering}

A short range magnetic field, $|B(x)|\leq C\langle x\rangle^{-1-\e}$,
does not substantially influence the non-magnetic scattering theory,
in particular asymptotic completeness holds. Long range potentials
can also  be included. The most general result is due to Robert
\cite{R}; previously Loss and Thaller treated the
$|B(x)|\leq C\langle x\rangle^{-3/2-\e}$ case in \cite{LT-87}
and they also considered the Dirac operator \cite{LT-88}.

The borderline case, when $B(x)$ decays as
$|x|^{-1}$ at large distances
is substantially more involved. In this case, there is no
decay on the vector potential. In the simplest $d=2$ dimensional, 
axially symmetric situation the lack of decay leads to
a dense point spectrum in the low energy region
while the spectrum is absolutely continuous above
an energy threshold \cite{MS}.
To study the scattering, for 
 simplicity one considers $d=2$
and assumes that the field is homogeneous of degree $-1$, i.e. in polar
coordinates $(r,\theta)$ it is given by
$B = b(\theta)/r$. 
 The analogous problem 
for non-magnetic scattering is the case of a homogeneous of degree zero
electric potential, $V(x)= U(x/|x|)$,  where the generic classical trajectories
are asymptotically straight and they select the directions of local
extrema of $U$. The classical trajectory in the borderline magnetic case
turns out to be a logarithmic spiral if the magnetic field has a definite sign.
If the total flux is zero, $\int_0^{2\pi} b(\theta)\rd \theta =0$, then
the trajectories are approximately  
straight lines (in the direction of the zeroes of $b(\theta)$);
if the total flux is non-vanishing but $b$ has no definite sign, then both
types of behavior may occur. The corresponding quantum scattering follows
these trajectories. Part of this picture has already  been proven in 
the recent work \cite{CHS}, the rest is work under preparation.

Scattering in constant magnetic field was first studied in 
\cite{AHS} where the asymptotic completeness of the one-particle scattering
for short range and Coulomb potential  was shown 
(a different proof given in \cite{S-79(c)}). The general long range
potential was treated independently by {\L}aba \cite{La-93, La-95} and Iwashita
\cite{Iw}.

\subsection{Miscellaneous}

In this section we mention two results
whose non-magnetic counterparts are classics
but their standard proofs are quite rigid and
their extensions to magnetic fields were considerably
more involved.
 
The Rayleigh-Faber-Krahn isoperimetric inequality on the lowest Dirichlet
eigenvalue of a domain of given area predates quantum
mechanics. Its magnetic analogue asserts  \cite{E-96(1)} that 
the lowest magnetic Dirichlet eigenvalue, $\lambda(\Omega, B)$,
of a planar domain $\Omega\subset\bR^2$ with a fixed area 
and with a fixed  homogeneous magnetic field $B=const$
 is attained exactly for the disk
\be
   \lambda(\Omega, B) \ge \lambda(D, B), \qquad \mbox{Area}(\Omega)
   =\mbox{Area}(D), \qquad D =\mbox{disk} \; .
\label{isop}
\ee
The constant magnetic field plays the 
role of the homogeneity of the membrane
in Rayleigh's original formulation of the problem.

In the non-magnetic case, the minimal
eigenvalue in any dimensions
is attained for  the  ball. The minimizing domain
for constant magnetic fields in dimensions $d\ge 3$,
however, is unknown. Isoperimetric results for the 
magnetic Neumann Laplacians are also unknown.
Note that (\ref{isop}) does not hold for the Neumann case
since the ground state has a tendency to favor
non spherically symmetric geometry
(Section \ref{sec:ind}),
but the disk geometry  should be extremal for other spectral
variational problems 
 in this case as well.
 
The standard proof of the original
Faber-Krahn inequality uses rearrangement methods
that are applicable for positive functions. The magnetic
ground state of a general domain is  genuinely complex 
and its amplitude, its phase, and 
the vector potential must be rearranged separately.

\medskip

A Schr\"odinger operator with a periodic external potential
has purely absolutely continuous (AC) spectrum by a classical
theorem of Thomas \cite{Thom}. The periodicity of the magnetic field
itself does not guarantee AC spectrum (e.g.  $B=const\neq 0$
in $d=2$), but a periodic vector potential does.
Note that this latter implies not only the periodicity of
the magnetic field but also that the flux is zero in 
the unit cells.

The absolute continuity of 
the magnetic Schr\"odinger spectrum with a small periodic
vector potential was first proven in \cite{HH}.
The proof was reduced by perturbation
to the original analyticity argument of Thomas
and it could not be extended beyond the perturbative regime.
In \cite{BS} a representation similar to (\ref{qh}) was used
to transform a periodic vector potential into a periodic
external potential and a modification of Thomas' argument
 applied. A  periodic metric can also be
included \cite{Mo}.
This approach, however, works only in $d=2$ dimensions.

The general case was obtained by Sobolev \cite{So-99}
who proved that the spectrum is purely AC for 
the magnetic Schr\"odinger operator with 
a sufficiently smooth periodic vector potential in any dimension.
The proof combined Thomas' argument with a pseudodifferential
technique.

\section{Quantitative properties of one-body operators}\label{sec:1}

\subsection{Lieb-Thirring inequalities}\label{sec:lt}

One of the fundamental results about the standard
Schr\"odinger operator $-\Delta +V$ is the Lieb-Thirring
bound \cite{LT} on the moments of negative eigenvalues, $E_j$,
in terms of integral norms of the negative parts of the potential, $V_-$
$$
 \tr\; [-\Delta + V]_-^\gamma =
 \sum_j |E_j|^\gamma \leq L_{d,\gamma} \int [V]_-^{d+\gamma/2} \;
$$
with a finite constant $L_{d,\gamma}$ for
$d\ge 3$, $\gamma\ge 0$; $d=2$, $\gamma>0$ or  $d=1$, $\gamma\ge 1/2$.
This bound plays a crucial role in the proof of the stability of matter
and it provides a basic apriori estimate for the semiclassical
formulas and for justification of the Thomas-Fermi
theory for the ground state energy of atoms and molecules.

By the diamagnetic inequality, the usual proof of 
the Lieb-Thirring (LT) bound 
for the non-magnetic operator,
 $-\Delta+V$, applies  directly  to the magnetic Schr\"odinger operator,
$(p+A)^2+V$ as well.  The same holds for 
the Cwickel-Lieb-Rozenblum  
(CLR) bound on the number of eigenvalues ($\gamma=0$).
The presence of a magnetic 
field should, in principle,
improve these estimates, but no such non-trivial result is available.

The systematic study of Lieb-Thirring bound and semiclassics
for  the Pauli operator has started with a series of seminal
papers by Lieb, Solovej and Yngvason \cite{LSY-I, LSY-II, LSY-III}. 
For the $d=3$ dimensional Pauli operator,  $H=[\bsigma\cdot(p+A)]^2 +V$,
with  a constant magnetic field, $B$,
and external potential, the following
bound was proven for the sum of negative eigenvalues of $H$ \cite{LSY-I}
\be
    \sum_j |E_j| \leq  \mbox{(const.)} \int [V]_-^{5/2}+
\mbox{(const.)} \int |B|\, [V]_-^{3/2} 
\label{mlt}
\ee
where $[V]_-= -\min \{ 0, V\}$.
A similar bound holds in $d=2$ dimensions as well \cite{LSY-III}.
The first term in (\ref{mlt}) is the corresponding Lieb-Thirring estimate
for $-\Delta+V$.
Due to the paramagnetism,  the Pauli energy may be
below the non-magnetic energy and the
additional term $\int |B| \, [V]_-^{3/2}$ is indeed necessary.
The number of eigenvalues can be infinite in $d=2,3$ dimensions, so there
is no CLR-bound for the Pauli operator.

For non-homogeneous magnetic fields, the bound (\ref{mlt}) does not hold.
Most importantly, the existence of the Loss-Yau zero modes
shows that, in the perturbative regime,
the lowest eigenvalue itself may scale linearly in $[V]_-$.
Moreover, the pointwise density of the Loss-Yau zero mode
scales as $\max |\psi(x)|^2\sim B^{3/2}$. Therefore a general LT
estimate in the strong field regime must contain a term
that grows as the 3/2 power of $B$.
 
To prove a LT estimate with the $B^{3/2}$ scaling,
the spin-coupling term $\bsigma\cdot B$
in (\ref{lich}) is treated as a potential and the diamagnetic inequality
is used for $(p+A)^2$. Several papers \cite{E-95, So-97, Sh-99, BFFGS}
used this  idea with different assumptions on the magnetic field.
The most general result in this direction is due to Lieb, Loss and Solovej
\cite{LLS} showing that
\be
    \sum_{j} |E_j| \leq \mbox{(const.)} \int [V]_-^{5/2}+
\mbox{(const.)} \| B\|_2^{3/2}\, \| [V]_-\|_{4} .
\label{mlt1}
\ee
The proof introduces the so-called {\it running energy scale method}.
It consists of artifically scaling down the Pauli kinetic energy
in an energy-dependent way to reduce the negative effect of $\bsigma\cdot B$.
The main advantage of this method is that it uses no other assumptions
on $B$ apart from the finiteness of its $L^2$-norm. Note that
$\int B^2$ is the field energy.

Although a term growing as $B^{3/2}$ (in the large field regime)
is necessary for a general LT bound, a smaller power 
is sufficient if some control on $\nabla B$ is allowed.
Especially, the linearity in $B$ of the bound (\ref{mlt})
reflects the basic fact that the space with a magnetic
field cannot be considered isotropic: 
the magnetic field affects only the quantum motion in
the transversal directions.

Under a control on the $H^1$ norm of $B$
the LT bound in \cite{BFG} scales as the 17/12 power of the field.
With more regularity on $B$ and $V$ the lower power 5/4
was obtained in \cite{ES-I}.
Finally, the correct linear behavior in the field strength
under a stronger regularity assumption was proved in 
\cite{ES-IV} and \cite{ES-V}. 
The proof in \cite{ES-IV} is shorter, but the estimate
is not local: a large irregular magnetic field
far away from the support of $[V]_-$ should not
influence the eigenvalue sum too much, but the estimate
in \cite{ES-IV} does not reflect this.
A conceptually different proof was given in \cite{ES-V}
that relies on a much stronger localization and approximation
technique.

The main difficulty behind these proofs 
is to control the density of Loss-Yau zero modes. 
It is amusing to note that it was a substantial
endeavour to show that zero modes may exist at all
(Section \ref{sec:zm}). On the other hand, it
is quite difficult to prove an upper bound
on their number in terms of the expected first
power of the magnetic field \cite{ES-V}.

\subsection{Semiclassics and strong fields}

We have seen that magnetic fields  can cause surprising effects
when the magnetic lengthscale is comparable with other
lengthscales in the problem. However, 
in the semiclassical and/or in the strong field
regimes, lengthscales are typically separated, rendering
simpler  formulas available in the limit.
One studies the magnetic Schr\"odinger or Pauli operators
with two parameters:
$$
   H(h,b):=  (hp+bA)^2 + V\;\; \; \mbox{or}\;\; \;
 [\bsigma\cdot (hp+bA)]^2 + V \; ,
$$
where $h\ll 1$ is the semiclassical parameter
and $b$ is the field strength; in most cases $b\gg 1$ (assuming
that $A$ and $V$ are fixed).
The magnetic field is $bB(x)=b\, \mbox{curl}\, A(x)$.
Under these scalings, the magnetic field can typically be
approximated by a (locally) homogeneous one, since the
magnetic lengthscale $(b/h)^{-1/2}$ is short.
If, in addition, $hb\ll 1$, then the gap between
(local) Landau levels shrinks to zero and
magnetic effects usually do not contribute to the main term
in the asymptotic regime. This is especially the case
for the standard semiclassics when $h\to0$ and $b$ is fixed
\cite{CSS}.
 If, on the other hand, $hb\not\to0$,
then the main term is typically obtained by 
replacing the magnetic field by a locally constant one.

The key technical step is, therefore, to localize the problem
to a sufficiently small scale, where the data
(especially the magnetic field) can be considered 
homogeneous. Similarly to the non-magnetic theories,
two basic techniques have been developed:  pseudodifferential calculus
and coherent states.

Note that the presence of a constant magnetic field changes
 structure of the phase space.
 For example, in three dimensions, the phase space is 
$\bigcup_{\nu=0}^\infty  \bR^4_\nu$, where $\nu$ labels the Landau
levels.  The phase space for each level is four dimensional; it
consists of three position coordinates and only
one momentum coordinate which represents the free motion
parallel with the magnetic field.
 The momenta transversal to the field
are not present due to the localization effect of the field.
Accordingly, one either has to develop a pseudodifferential
calculus that treats the harmonic oscillators on each Landau
level exactly or one has to construct magnetic coherent states.

\medskip

\subsubsection{Results on individual eigenvalues}
\label{sec:ind}

Semiclassical estimates on individual eigenvalues and eigenfunctions
have mainly been carried out for the ground states.
For the magnetic Schr\"odinger operator without potential
 the ground state is localized near the minimum of the field strength
(``magnetic bottles'', see \cite{AHS} and \cite{CdV}).
The basic observation is that due to the positivity of the
Pauli operator and the Lichnerowicz formula (\ref{lich}),
 the magnetic Schr\"odinger operator
can always be estimated from below by $B(x)$, 
 at least in two dimensions and if $B\ge 0$.
Similar result holds in higher dimensions, at least locally.
This concentration of ground state is especially visible
in the large field regime. Note that if external potential is not present, 
the semiclassical limit is formally identical
to the large magnetic field limit, $h=1/B$.

 Precise analysis of this  phenomenon
was initiated by several groups with different methods.
With the help of the Feynman-Kac formula, the magnetic ground
state energy can be turned into a question about the
rate of decay of an oscillatory Wiener integral, see
\cite{Ma-86} and for a more precise bound \cite{E-94}, \cite{U-94}.
Ueki has also explored the connection with the hypoellipticity
of the $\overline{\partial_b}$ problem 
 \cite{U-02}. Montgomery \cite{Mon} 
has analysed the case when the two dimensional magnetic field vanished
along a curve. In this case $\min |B|=0$, hence
the leading term in the large field asymptotics vanishes,
and Montgomery obtained the subleading term that involved
the curvature of the zero locus. This
approach was later generalized 
in \cite{HM-96}.

On a domain with boundary,  however, the Lichnerowicz formula (\ref{lich})
does not hold, unless Dirichlet boundary condition are imposed.
In particular, the ground state energy of the magnetic
Schr\"odinger operator with Neumann boundary condition
is smaller than $B$ even for a constant magnetic field. In this latter case
the ground state is localized near the boundary, more precisely
near the point with largest curvature of the boundary.
The second term in the semiclassical expansion of the
ground state energy is determined by the curvature, similarly
to Montgomery's result.
Similar phenomenon occurs in three dimensions as well
with a proper definition of an effective curvature \cite{HM-04}.
Recently a complete expansion for the energy of the low
lying eigenvalues in $d=2$ dimensions was carried out in
\cite{FH}. We remark that 
the Neumann boundary problem naturally arises
at the minimization of the Ginzburg Landau energy
functional describing superconducting states.
Several people have contributed to these results, see 
\cite{HM-04}, \cite{FH} and references therein.

\subsubsection{Results on cumulative spectral quantities}

The first result on spectral statistics, where the magnetic field
played a non-trivial role, is probably due to Colin de Verdi\`ere
\cite{CdV} and  Tamura \cite{Tam} (independently)
who proved a Weyl-type asymptotics for the
number of eigenvalues with a non-homogeneous
magnetic field and a confining potential. The magnetic field increases
at infinity, ensuring that magnetic effects contribute
to the large energy asymptotics.
Colin de Verdi\`ere used
the magnetic extension of the classical Dirichlet-Neumann bracketing,
while Tamura  estimated the short time asymptotics of the magnetic heat
kernel. Several authors extended these results, see \cite{MU} for references.

A technically somewhat similar problem is the
rate of the eigenvalue accumulation near the Landau
levels due to perturbation by a decaying potential.
The basic idea is that the magnetic field
strongly localizes the particle  and its
interaction with the potential can be computed 
fairly precisely. The accumulation rate
is explicitly given by the decay rate of the
 external field, see \cite{RW} and references therein.

\medskip

The next semiclassical question concerns the moments of negative eigenvalues
in the spirit of \cite{LS}.
Here we consider only the more interesting case 
of the Pauli operator, $H(h, b):= [\bsigma\cdot (hp+bA)]^2 + V $. 
For simplicity, we work
in $d=3$ dimensions and we will approximate
only the eigenvalue sum;
$$
     \Sigma(h, b): = \tr \big[ H(h,b)\big]_- \; .
$$
The corresponding semiclassical expression is
$$
 E_{sc}(h,b): = -h^{-3}\int_{\bR^3} P(hb|B(x)|, [V(x)]_-)\rd x
$$
with
\be
  P(B, W):= \frac{B}{3\pi^2}\Big( W^{3/2} + \sum_{\nu=1}^\infty
 [ 2\nu B -W]_-^{3/2}\Big)  \; .
\label{pres}
\ee
This formula
 can be simply 
deduced from the structure of the phase space outlined above.

For homogeneous magnetic field, the semiclassical limit
\be
    \lim_{h\to 0} \frac{\Sigma(h,b)}{E_{sc}(h,b)}=1
\label{sc}
\ee
was proved uniformly in the field strength $b$
\cite{LSY-II} (the two-dimensional
result was obtained in \cite{LSY-III}). 
The main ingredients were the magnetic
Lieb-Thirring inequality (\ref{mlt}) and new magnetic coherent states.
For the non-homogeneous case, a Lieb-Thirring inequality
that scales as $B^{3/2}$ (see Section \ref{sec:lt}) allows one
to prove the semiclassical formula only up to $hb = O(1)$
\cite{So-98}. With the improved Lieb-Thirring inequality 
\cite{ES-I} and a new construction of
coherent states, the proof can be extended to $b\ll h^{-3}$.
This result is already  sufficient to cover the full semiclassical
regime of the large atoms \cite{ES-II}, see Section \ref{sec:large}.
Uniform semiclassics can be obtained with the help of
the uniform Lieb-Thirring inequalities \cite{ES-IV, ES-V}.

The development using pseudodifferential calculus has focused
on obtaining precise spectral asymptotics for the local
traces of the form $\tr \, [\,\chi \varphi(H)\, ]$, where $\chi$
is a spatial cutoff function.  These efforts have culminated
in the book of Ivrii \cite{I} 
 where  precise remainder estimates were proven in a great generality.
His  recent work investigates the same questions  with 
irregular data \cite{I-05}.  A more concise
result using these ideas is the proof of a certain local version 
of (\ref{sc}) for homogeneous field
in \cite{So-94}, improved later in \cite{So-96}
to include  Coulomb singularities. The microlocal technique
gives also higher order corrections to the leading term. However,
these methods require, in general, strong regularity assumptions
on the data. Moreover, some non-asymptotic
apriori estimate (Lieb-Thirring bound) is necessary to remove
the cutoff $\chi$.

\medskip

In addition to the energy, other physical quantities are also of
interest. Fournais has studied the quantum current in  a
magnetic field and proved the corresponding semiclassical
formula. Note that the current is
a second order effect in the semiclassical expansion and it
becomes a leading term only after non-trivial cancellations.
Both microlocal techniques similar to \cite{So-94}
and coherent states methods similar to \cite{LSY-II}
have been tested, e.g. in \cite{F1} and \cite{F2}.

\subsection{Peierls substitution and corrections to the semiclassics}

The Bloch decomposition for a single particle Schr\"odinger
operator with a  periodic external potential can be 
extended to include weak electromagnetic fields. The basic
idea due to Peierls is to substitute the minimally coupled
magnetic momentum $p+A$ into the band functions, $E_n(p)$,
obtained from the non-magnetic Bloch decomposition.
If the electromagnetic field varies on a much larger
scale than the periodic background, then the problem is
effectively semiclassical with a
scale-separation parameter $\e$ (for the general theory, see \cite{N}).
The  resulting pseudo-differential operator can be
analyzed with well developed mathematical tools.
Algebraic methods were applied in \cite{BR}, for a systematic
presentation of the pseudo-differential  approach see \cite{HS-2}.
For example, it was shown in \cite{GMS}
that near a fixed energy level the original Hamiltonian is isospectral to 
a pseudodifferential operator with the same principal symbol
as the Peierls Hamiltonian has.
The detailed behavior of the density of states,
in particular the de Haas-van Alphen effect for
the oscillation of the magnetization was shown
in \cite{HS-3}.

The de Haas-van Alphen oscillation is due to a subleading effect 
in a semiclassical type expansion, however, it 
determines the current to leading order. 
The electromagnetic field is weak, but on the long time scale, $t\sim 
\e^{-1}$, it yields an order one change in the
dynamics.  
To describe these effects on the dynamics correctly, 
Panati, Spohn and Teufel \cite{PST}
have  developed a time dependent version of the Peierls substitution.
The classical equations are corrected by
an  the effective magnetic moment (Rammal-Wilkinson term)
and an ``anomalous velocity'' term due 
to the curvature of the Berry connection.
This latter,  in particular, provides
a simple semiclassical explanation
of the quantum Hall current.

We remark that an oscillation similar to the de Haas-van Alphen effect
is exhibited for the Harper operator (\ref{harper})
for magnetic fluxes (per unit cells) that are  near
a rational number, see \cite{GA} and references therein.
The Harper operator itself
can also be viewed  as a Peierls substitution by quantizing the classical
symbol $H(\xi, \eta)=\cos \xi +\cos \eta$ with $[\xi,\eta]= 2\pi i \alpha$.
In this case the distance of $\alpha$ to a nearby rational
number with small denominator plays the role of the semiclassical
parameter \cite{HS-1}.

\section{Many-body  magnetic systems}\label{sec:mult}

\subsection{Magnetic stability of matter}

In a system of charged fermionic quantum particles subject to a Coulomb
interaction the ground state energy per particle is uniformly bounded,
independently on the number of particles. This fundamental fact is called
the stability of matter. For an excellent review of the progress in
the last 35 years, see \cite{L-04}.

The first proof of the stability of matter with
the nonrelativistic kinetic energy, $-\Delta = p^2$,
is due to Dyson and Lenard.
A simpler proof was given later  by Lieb and Thirring
using the Lieb-Thirring inequality.
 Stability of matter also
holds if the nonrelativistic
kinetic energy operator, $p^2$, is replaced by  $|p|=\sqrt{-\Delta}$ 
(``relativistic'' kinetic energy) and   the fine
structure constant, $\alpha$, is sufficiently small.
The first proof was given by Conlon, improved by Fefferman-de la Llave
and finally the optimal bound was obtained by Lieb and Yau; see
also a recent improvement and references in \cite{LLSi}.

In the relativistic case, the kinetic energy and the potential energy
both scale as $[length]^{-1}$, therefore the energy per particle
can be bounded from below only if the Hamiltonian is non-negative, i.e.
\be
     \sum_{j=1}^N |p_j| + \alpha V_c \ge 0
\label{liy}
\ee
where $V_c$ stands for the Coulomb potential of $N$ electrons and $K$ nuclei
with charges $Z$.
This inequality was proven by Lieb and Yau
in \cite{LYa} (Theorem 2) for $\alpha \leq 1/94$
and $Z\alpha \leq 2/\pi$, the second condition being optimal.
Theorem 1 of \cite{LYa} has a weaker result ($\alpha \leq 0.016$,
$Z\alpha\leq 1/\pi$) but its proof can easily be generalized
to the magnetic case as well, where
 the kinetic energy $|p|$ can is
replaced by its magnetic counterpart $|p+A|$.
This follows from a simple application of the diamagnetic
inequality as it was pointed out in \cite{LLS}.

With the Pauli kinetic energy, however, even the hydrogen atom
is unstable because in a strong magnetic field
the electron can be strongly localized around the nucleus
without a penalty in the kinetic energy. The ground state
energy of the hydrogen in a constant magnetic field $B$
diverges as $(\log B)^2$, if $B$ is large \cite{AHS}.

If the energy of the  magnetic field is added to the total energy,
then stability is restored:
\be
   \sum_{j=1}^N \; [\bsigma\cdot(p+A)_j]^2 + V_c
 + \frac{1}{8\pi\alpha^2}\int B^2
  \ge - C(Z) (K+N)
\label{mst}
\ee
where the constant depends only on the charges of the nuclei.
The parameter $\alpha$ (fine structure constant) must be sufficiently small
in order (\ref{mst}) to hold.
The existence of a Loss-Yau zero mode \cite{LY} shows that
the total energy can be negative if $\alpha$ is large.
Actually, an absolute upper bound on $\alpha$ and
an upper bound on $(\max Z_j)\alpha^2$ are both necessary,
where $Z_j$ are the nuclear charges.

The bound (\ref{mst})  is called the stability of matter interacting with
a classical magnetic field. It was first proven for atoms \cite{FLL}
and single electron molecules \cite{LL-86}.
The general case was
settled  by Fefferman \cite{Fef} for a very small
$\alpha$.    Lieb, Loss and Solovej
\cite{LLS} gave a much shorter proof that also  holds 
for the physical value of the fine structure constant.
The backbone of this proof is the magnetic Lieb-Thirring
inequality (\ref{mlt1}).

Using the Birman-Koplienko-Solomyak trace
inequality and the magnetic version of the Lieb-Yau bound (\ref{liy}), 
Lieb,  Siedentop and Solovej \cite{LSS} also proved the stability 
of matter for the Dirac operator. The particles
must be restricted to the
 positive energy subspace of the
one-particle Dirac operator $\balpha \cdot (p+A) +\beta$ 
(filling the ``Dirac sea''). It is important to note
that the Dirac sea must be defined  via the gauge
invariant Dirac operator. Restriction onto the positive
energy subspace of the free Dirac operator (as it
is often done in perturbation theory) leads to 
instability of matter for any $\alpha$.

Ultimately, the electromagnetic field must also be quantized.
Imposing and ultraviolet cutoff and using results from
\cite{LLS}, Bugliaro, Fr\"ohlich and Graf proved stability
of matter for the Pauli operator
with a quantized electromagnetic field \cite{BFrG}.
The essential observation is that to restore the magnetic
stability for the Pauli operator with a classical field,
it is sufficient to add the field energy only near the nuclei.
On a finite volume and with an ultraviolet cutoff,
the classical and quantized field energies can be compared.
Lieb and Loss \cite{LL-02} showed stability of matter
for the Dirac operator with a quantized electromagnetic field
and with a suitable one-body spectral projection, similar to \cite{LSS}.

The quantization of the electromagnetic field poses several 
complications, such as ultraviolet cutoff and mass renormalization,
and little is known about how to rigorously include them into a 
fully consistent theory.
However, one problem has been settled satisfactorily: the existence of
atoms in nonrelativistic QED.  Because
of  the quantized field, the ground state of the total system (atom
and photons) is 
at the bottom of a continuous spectrum, and it is not at all
obvious that it is an eigenvalue. Moreover, the so-called
binding condition also has  to be satisfied, that is
the energy of a system of $N$ electrons is actually lower
than that of a system with fewer electrons, otherwise the
ground state may contain no electron at all.
For small values of the parameters (ultraviolet cutoff
parameter and fine structure constant) the existence of
atoms was shown in \cite{BFS} and 
for arbitrary parameter values in
\cite{GLL, LL-03}. No infrared cutoff was needed, unlike
for the scattering problem (Section \ref{sec:scat}).
More recently, the thermodynamic limit
for non-relativistic Coulomb matter with quantized electromagnetic
field was investigated and the lower bound  was proved
\cite{LL-05}.

\subsection{Large atoms}\label{sec:large}

One of the main motivations to study semiclassical spectral
asymptotics  for $-h^2\Delta +V$ originates in the
seminal paper of Lieb and Simon, \cite{LS}, where 
the exactness of the Thomas-Fermi theory for
the ground state energy of atoms with large nuclear charge, $Z\gg 1$,
was proven with semiclassical methods. 

In the presence of magnetic fields the Thomas-Fermi theory
is more complex. Depending on the strength of the magnetic
field compared with $Z$, there are five different
regimes.

Within the classical Thomas-Fermi theory, 
the kinetic energy as a functional of the density is always given by
the Legendre transform of the pressure (\ref{pres}).
This classical magnetic Thomas-Fermi  theory, however,  holds only for weak
and moderate magnetic fields ($B\ll Z^3$). More precisely,
 for $B\ll Z^{4/3}$ the magnetic effects are absent
in the leading term of the large $Z$ asymptotics.
For $B\sim Z^{4/3}$ the full magnetic Thomas-Fermi theory is needed.
If $Z^{4/3}\ll B\ll Z^3$, the usual Thomas-Fermi theory
still applies, but only the first summand in the pressure
function (\ref{pres}) is needed. The atom is  spherical
in all these cases, but the energy is affected by the
magnetic field. These results  were
proven in \cite{LSY-II} with the help of the magnetic Lieb-Thirring
inequality and the semiclassical result (\ref{sc}).

 For a strong magnetic field,
$B\ge (const) Z^3$ with a positive constant,
 the electrons are confined to the lowest
Landau band and the shape of the atom is a long cylinder
along the field. For $B\sim Z^3$ the atoms in the
 transversal direction  have a non-trivial structure,
that can be described by a new density functional theory 
relying on minimizing density matrices instead of density
functions \cite{LSY-I}. Finally, if $B\gg Z^3$, the
atom becomes effectively one-dimensional and a
one dimensional Thomas-Fermi caricature applies.
 Analogous results in $d=2$
were obtained in \cite{LSY-III}.

If the magnetic field goes to infinity, but $Z$ is fixed,
then the ground state energy diverges as $-\frac{1}{4} Z^2 (\log B/2)^2$.
For one electron atoms this has been established in \cite{AHS}.
The energy of the many-body system, after
factoring out the divergent $(\log B)^2$ term,
is given by the ground state energy
of an effective one dimensional {\it bosonic} 
Hamiltonian with Dirac delta interactions \cite{BSY}.
This idea has been extended to prove resolvent convergence 
and explore other effective Hamiltonians
in \cite{BD} and references therein.

The correctness of the Thomas-Fermi theory in
 the semiclassical regime ($B\ll Z^3$) 
for non-homogeneous magnetic fields were proven in
\cite{ES-II} after extending the Lieb-Thirring
inequality \cite{ES-I} and constructing
appropriate coherent states. The uniform magnetic Lieb-Thirring
inequality \cite{ES-IV} should allow also the extension
of the strong field regime from \cite{LSY-I} to
non-homogeneous magnetic fields.

The asymptotic behavior of the  total magnetization and of the current
for large atoms in homogeneous fields was obtained in \cite{F2, F3}.
A bound on the maximal ionization is proven in \cite{S}.

\subsection{Multiparticle scattering in a magnetic field}\label{sec:scat}

A detailed presentation of scattering theory and asymptotic
completeness is given in 
the contribution of C. G\'erard in this volume; here we just
shortly mention the most important results involving magnetic fields.

$N$-body asymptotic completeness is well understood for
non-relativistic particles without magnetic field.
Scattering in the presence of a constant magnetic field is a much more
delicate question since a classical charged particle moves on circles.
Therefore charged subsystems can scatter only parallel with the
field, while neutral systems may move out to infinity in all directions.
The general theory has been developed and asymptotic completeness has been
proved for the case when all
possible subsystems are charged by G\'erard and {\L}aba
(the best reference is their book \cite{GL}).
The zero charge case in general is still open.
The  special case of three particles with Coulomb forces
was solved  in \cite{GL1} and one charged particle was considered 
in \cite{Ad}.

Constant electric field
can also be included. Skibsted \cite{Sk}
reduced this problem to scattering 
of non-interacting subsystems all having the same charge/mass ratio.

In the presence of a quantized electromagnetic field, the scattering
of photons on an bound electron (Rayleigh scattering) and
the scattering of electron dressed with photons (Compton scattering)
have been studied. The mathematical framework is non-relativistic
quantum mechanics with a quantized field with ultraviolet cutoff
to ensure that the Hamiltonian is well defined.
For total energies below the electron-positron pair creation
threshold, this non-relativistic caricature of QED is
well justified. One also has to cope with the infrared (IR)
problem (``soft bosons''): there could be infinitely
many photons with a finite total energy.  For 
technical simplicity,  scalar photons  are considered.
This set of problems was initiated by  Fr\"ohlich
who first investigated the infrared problem and constructed
wave operators (without completeness) \cite{Fr}.

The Rayleigh scattering was first tackled in \cite{DG}
where the photons were massive to avoid the IR problem
and the potential was confining.
Later it was extended to the physically more realistic
 massless photons, but with an
IR-cutoff, and instead of confining potential, the total
energy was set below the ionization threshold which also
guarantees spatial localization of the electrons
\cite{FGS1}.  During scattering, photons are
absorbed by the bound electrons, lifting them to higher
excited states, but no electron can escape.
Thus, after some time, the electron cloud relaxes to the ground state
and emits photons that propagate essentially freely
to infinity in space. Note that the  analysis of
Bach et al. guarantees that all excited states are unstable 
\cite{BFS, BFSS}.

For Compton scattering  \cite{FGS2}, the total energy is assumed to be
sufficiently small  so that the speed of the massive electron 
is less than 1/3 of the
 speed of light. This technical  assumption safely separates free photons
from the dressed electron after long time evolution.
The asymptotic completeness in this model means that
the long time evolution of the state is a linear combination
of asymptotic states consisting of a freely moving electron dressed
by a photon cloud plus freely moving excess photons.

\section{Random Schr\"odinger operators with magnetic fields}
\label{sec:random}

Since the proof of the Anderson localization 
with the powerful multiscale method of Fr\"ohlich and Spencer
\cite{FS}, random Schr\"odinger operators have become
one of the main research directions in mathematical physics.
The typical problems concern the self-averaging properties
(deterministic spectrum, existence of density of states),
the asymptotics of the density of states (Lifshitz tail),
and establishing the dense point spectrum in the
localization regime. These
questions have been recently studied with magnetic field as well;
most results are in the most relevant $d=2$ dimensions.

\subsection{Constant magnetic field}

First we consider the problems where
 the magnetic field is constant and only the external
potential is random, i.e. the random Landau
Hamiltonian (for a recent survey, see \cite{LMW}
and references therein). In this case, 
the standard self-averaging techniques work to establish
deterministic spectrum, the translation 
must simply be replaced by the magnetic translation.
The Lifshitz tail in the low energy regime for Gaussian
random potential is very universal and it is insensitive to
the magnetic field (it even holds for certain random magnetic fields).
For Gaussian randomness even the density of states is bounded
and the Wegner estimate holds \cite{HLMW}. Localization 
with algebraically decaying eigenfunctions were proved
in \cite{FLM}, exponential localization in \cite{U-02}. 
These proofs are valid only at very low energies, using
the deep wells of the Gaussian randomness, in particular 
this regime is far from the analogue of the band-edge
localization.

For repulsive Poissonian obstacles the precise Lifshitz tail is
more delicate, similarly to the classical vs. quantum
dichotomy in the non-magnetic setup.
If the single-site potential has a slow decay, then
classical effects dominate and the Lifshitz tail
can be computed from a simple mean-field argument.
Otherwise the quantum localization energy competes
with the entropy of the large domains free of impurities.
For magnetic fields, the threshold decay is 
not algebraic but Gaussian. 
The classical regime was investigated in \cite{BHKL}
(in three dimensions \cite{HKW}), the main result in the
quantum regime was obtained in \cite{E-98} using Sznitman's
coarse-graining probabilistic method (see also 
\cite{HLW}, \cite{E-01}).

Since in the Poisson model the energy is not monotone
in the random variable, so Wegner estimates are
much harder to obtain, localization for a constant magnetic
fields were investigated for Anderson type i.i.d. random potentials.
Wang has given a full asymptotic expansion
of the density of states away from the Landau bands \cite{W-00}.
More explicit quantitative results are known on the Lifshitz tails,
including the double logarithmic asymptotics at
each band edge \cite{KR}.
The pure point spectrum at a certain distance away from the Landau levels
was proven independently by Wang \cite{W-97}, Combes and Hislop
\cite{CH} and in a single-band approximation by Dorlas, Macris
and Pul\'e \cite{DMP}.
Since $d=2$ is the borderline dimension
for the necessary large distance decay of 
the Green's function of the non-magnetic Laplacian,
an additional decay  must be extracted from the
presence of the magnetic field. This eventually required
basic results from two-dimensional bond 
 percolation theory. Precise
estimates on the localization length and dynamical localization near
the Landau band edges  was obtained by Germinet and Klein \cite{GK-03}
using their extension of the Fr\"ohlich-Spencer  multiscale analysis
\cite{GK-01}.

In contrast to the localization regime, the presumed
regime of delocalization in  Anderson-type models
is poorly understood and
there are almost no mathematical results. Apart
from the Bethe-lattice, there is only one model,
where the existence of the mobility edge
has been rigorously proven: the random Landau
Hamiltonian \cite{GKS}. More precisely, it is
shown that there exists at least one energy $E$ near
each Landau band, so that the the local transport
exponent, $\beta(E)$, is positive. The local transport
exponent measures the extension of the wave packet
in a suitable averaged sense for large times.  (Dynamical) localization
is characterized by $\beta(E)=0$, moreover, there is an important
dichotomy for $\beta(E)$: it is either zero or at least $1/2d$.

The key quantity in the proof in \cite{GKS} is the Hall conductance.
In the regime of dynamical
localization, the Hall conductance is constant in the mobility gap 
(see \cite{BES}, strengthened later in \cite{EGS}).
On the other hand, the Hall conductivity jumps by one at each
Landau level for the free Landau Hamiltonian and it is also
 constant, as a function of the disorder parameter,
in the gaps. Therefore it must jump 
somewhere inside the bands, at least for sufficiently small disorder,
but then the complete band cannot belong to the dynamically localized
regime, completing the argument of delocalization 
in \cite{GKS}.

The Hall conductance for  quantum Hall systems 
(two dimensional disordered samples
subject to a constant perpendicular magnetic field)
at energies falling into the mobility gap, $\Delta$,
can actually be defined in two different ways.
The bulk conductance is defined on $\bR^2$ by the Kubo-Streda
formula (see \cite{ASS})
$$
    \sigma_B(\lambda) = -i\tr P_\lambda \; [\; [P_\lambda, 
\Lambda_1], [P_\lambda, \Lambda_2]\; ]
$$
where $P_\lambda$ is the spectral projection onto $(-\infty, \lambda]$
and $\Lambda_i$ is the characteristic function of $\{ x_i < 0\}$,
$i=1,2$.

The edge conductance is defined in a half plane sample
($x_2\ge -a$, eventually $a\to \infty$) by
$$
   \sigma_E = -i\tr \varrho'(H) [H, \Lambda_1]
$$
(modulo nontrivial technicalities),
where $\varrho$ is a smooth spectral cutoff  function
that is one below $\Delta$ and zero above $\Delta$.
This quantity gives the derivative of the current flowing
through the line $x_1=0$ with respect to 
lowering the chemical potential along the edge.
These two conductances are the same. For intervals $\Delta$
falling into the spectral gap, this was proven in
\cite{SKR} with K-theoretical methods 
and later in \cite{EG} by basic functional
analysis.  The proof given recently in \cite{EGS} is valid
for intervals $\Delta$ that contain strongly localized
spectrum as well, i.e. $\Delta$ can be chosen 
the so-called mobility gap.

\subsection{Random magnetic field}

Since a magnetic field itself enhances localization, one
expects that a random magnetic field is localizing even stronger
than a random potential. Technically, however,
random magnetic fields are harder to fit into
the multiscale analysis; mainly because the
vector potential is non-local. 
There is a substantial difference between the zero and non-zero
flux cases, the former being easier.
In particular, stationary random vector potentials
always generate magnetic fields that have zero flux on average.

The existence  of the integrated density of states (IDS),
and its independence of the boundary conditions in the thermodynamic
limit (uniqueness) has first been shown by Nakamura
for both the discrete and 
continuous Schr\"odinger operator 
with a random magnetic field with non-zero flux
\cite{Na2, Na3}. In both cases Lifshitz tail was also obtained.
Recently, Hundertmark and Simon gave a short proof for
the existence and uniqueness of the IDS \cite{HS}.

Anderson localization has been shown for a Gaussian
vector potential  by Ueki in \cite{U-02}. The Germinet-Klein
multiscale analysis has been extended to include very
general random magnetic fields, but the Wegner estimate
requires random vector potential (which implies zero average flux)
in addition to other technical conditions.

For the discrete random magnetic Schr\"odinger operator,
the method of Nakamura \cite{Na2} has been extended 
to obtain Wegner estimate and localization \cite{KNNN}. However, the zero flux
condition is enforced in a strong sense: neighboring cells
are paired and the magnetic flux is opposite in these pairs.
A deterministic background magnetic field atop
of the local random vector potential is allowed 
in \cite{HK} where Wegner estimate was proved in the continuous model.
A small stationary random vector-potential was included 
in a Schr\"odinger operator with a periodic background
potential in \cite{G}, where Lifshitz tail was proven under
a special non-resonance condition.

Band-edge localization for  Schr\"odinger operators
with random magnetic field is widely believed to hold in the most
general case. The additional assumptions on the zero  flux
(which, in one form or another, is present in all papers so far)
seems to be only technical. However, 
there is no agreement in the physics literature about
the possible existence of the continuous spectrum for such operators,
unlike in the non-magnetic case, where the existence of the  extended
states is universally accepted by physicists and ``only''
the mathematical proof is missing. The Landau orbits and their
quantum counterparts, the strongly localized magnetic eigenfunctions
are characteristic only to the constant magnetic field and they are 
not universal. There is an undecided competition between 
a possible weaker form of the  Landau localization,
that may still hold for random fields,
 and the  resonance effects that enhance delocalization. 

\bigskip

{\it Acknowledgement.} The author is indebted to Jan Philip Solovej
for his help in preparing the manuscript.  Special thanks to
Christian G\'erard, Gian-Michele Graf,
Dirk Hundertmark, Fr\'ed\'eric Klopp,  Michael Loss,
Benjamin Schlein, Stefan Teufel, Jakob Yngvason, Simone Warzel
and to the referee for their
critical reading and corrections.
Part of this work has been done at Harvard University.

\small{

}

\bigskip
\noindent Address of the author:

\medskip

\noindent  L\'aszl\'o Erd\H os \\
Mathematisches Institut, LMU\\
Theresienstrasse 39, D-80333 Munich, Germany\\
 lerdos@mathematik.uni-muenchen.de

\bigskip


\begin{thebibliography}{hhh}

\bibitem[1]{Ad} T. Adachi: {\em On spectral and scattering
theory for $N$-body Schr\"odinger operators in a constant
magnetic field.}  Rev. Math. Phys. {\bf 14}, 199--240 (2002)

\bibitem[2]{AC} Aharonov, Y. and  Casher, A.: {\em Ground state of spin-1/2
 charged particle in a two-dimensional magnetic field.} Phys. Rev.
 {\bf A19},  2461--2462 (1979).


\bibitem[3]{AJ} A. Avila, S. Jitomirskaya: {\em The ten Martini problem.}
Preprint 2005, {\tt xxx.lanl.gov/math.DS/0503363}

\bibitem[4]{ASS} J. Avron, R. Seiler, B. Simon:
{\em Charge deficiency, charge transport and comparison
of dimensions.} Commun. Math. Phys. {\bf 159}, 399--422 (1994)

\bibitem[5]{AS} J. Avron, B. Simon: {\em A counterexample
to the paramagnetic conjecture. \/} Phys. Lett. {\bf A75}, 41--42 (1979)

\bibitem[6]{AHS} J. Avron, I. Herbst and B. Simon: {\em Schr\"odinger
operators with magnetic fields. I. General interactions. \/} Duke Math. 
J. {\bf 45} (1978), 847--883.  {\em II. Separation of the center mass
in homogeneous magnetic fields. \/} Ann. Phys. {\bf 114}, 431--451 (1978),
{\em III. Atoms in homogeneous magnetic fields. \/} Commun. Math. Phys.
{\bf 79}, 529--572 (1981)



\bibitem[7]{AMN}  C. Adam, B. Muratori and C. Nash:
{\sl Multiple zero modes of the Dirac operator in three dimensions},
Phys. Rev. D (3) {\bf 62}, no. 8, 085026 (2000)

\bibitem[8]{BFS} V. Bach, J. Fr\"ohlich, I. M. Sigal: {\sl
Spectral analysis for systems of atoms and molecules
coupled to the quantized radiation field.}
Commun. Math. Phys. {\bf 207}, 249--290 (1999).

\bibitem[9]{BFSS}  V. Bach, J. Fr\"ohlich, I. M. Sigal, A. Soffer:
 {\sl
Positive commutators and the spectrum of Pauli-Fierz Hamiltonian
of atoms and molecules.} 
Commun. Math. Phys. {\bf 207}, 557--587 (1999)



\bibitem[10]{BE} A. Balinsky, W. D. Evans: {\sl
On the zero modes of Pauli operators}, J. Funct. Anal. 
{\bf 179}(1), 120--135 (2001)

\bibitem[11]{BES} J. Bellissard, A. van Elst, H. Schulz-Baldes:
{\em The noncommutative geometry of the quantum Hall effect.}
J. Math. Phys. {\bf 35}, 5373--5451 (1994)


\bibitem[12]{BSY} B. Baumgartner, J.P. Solovej and
J. Yngvason: {\sl Atoms in strong magnetic fields: the high
field lmit at fixed nuclear charge.}
Commun. Math. Phys. {\bf 212}, 703-724 (2000)


\bibitem[13]{BeS} J. Bellissard, B. Simon: {\sl Cantor spectrum
for the almost Mathieu equation.} J. Funct. Anal.  {\bf 48}, 408-423
(1982)


\bibitem[14]{BR} J. Bellissard, R. Rammal: 
{\sl An algebraic semi-classical approach to Bloch electrons
in a magnetic field.} J. Physique France, {\bf 51}, 1803 (1990)


\bibitem[15]{BS} M. Sh. Birman, T. A. Suslina:
{\em The two-dimensional periodic magnetic Hamiltonian is
   absolutely continuous.} St. Petersburg Math. J. {\bf 9},  21--32
(1998)




\bibitem[16]{BHL} K. Broderix, D. Hundertmark, H. Leschke:
{\em Continuity properties of Schr\"odinger semigroups
with magnetic fields.} Rev. Math. Phys. {\bf 12}, 181--225
(2000)


\bibitem[17]{BHKL} 
K. Broderix, D. Hundertmark, W. Kirsch, H. Leschke:
{\em The fate of Lifshits tails in magnetic fields.}
J. Stat. Phys. {\bf 80} 1--22 (1995)


\bibitem[18]{BD} R. Brummelhuis, P. Duclos:
{\sl Effective Hamiltonians for atoms in very strong magnetic
fields.}  {\tt xxx.lanl.gov/math-ph/0502050}



\bibitem[19]{BFFGS} L. Bugliaro, C. Fefferman, J. Fr\"ohlich,
G. M. Graf and J. Stubbe: {\sl A Lieb-Thirring bound for a magnetic
Pauli Hamiltonian}, Commum. Math. Phys. {\bf 187}, 567--582 (1997)


\bibitem[20]{BFrG} L. Bugliaro, J. Fr\"ohlich, and G. M. Graf:
{\sl Stability of quantum electrodynamics with nonrelativistic matter},
Phys. Rev. Lett. {\bf 77}, 3494--3497 (1996)

\bibitem[21]{BFG} L. Bugliaro, C. Fefferman and G. M. Graf:
{\sl A Lieb-Thirring bound for a magnetic
Pauli Hamiltonian, II}, Rev. Mat. Iberoamericana, {\bf 15}, 593--619
(1999)


\bibitem[22]{CdV}  Y. Colin de Verdi\`ere:
 {\em  L'asymptotique de Weyl pour
les bouteilles magn\'etiques. \/} Commun. Math. Phys. {\bf 105},
327--335 (1986)




\bibitem[23]{CdVT} Y. Colin de Verdi\`ere, N. Torki:
{\em Op\'erateurs de Schr\"odinger avec champs magn\'etique.}
 S\'eminaire de th\'eorie spectrale et g\'eom\'etrie (Grenoble)
{\bf 11}, 9--18 (1992-1993)

\bibitem[24]{CH} J. M. Combes, P. Hislop: {\em Landau Hamiltonians
with random potentials: localization and the
density of states.} Commun. Math. Phys. {\bf 177}, 603--629 (1996)

\bibitem[25]{CSS} J. M. Combes, R. Schrader, R. Seiler:
{\em Classical bounds and limits for energy distributions
of hamiltonian operators in electromagnetic fields.}
Ann. Physics {\bf 111}, 1--18 (1978)

\bibitem[26]{CHS} H. Cornean, I. Herbst, E. Skibsted:
{\em Spiraling attractors and quantum dynamics for a class of long-range
magnetic fields.} Preprint.  Univ. Texas Mathematical
Physics Archive, 02-452 (2002)

\bibitem[27]{CFKS} H. L. Cycon, R. G. Froese, W. Kirsch and B. Simon:
{\em Schr\"odinger Operators with Application to Quantum Mechanics and
Global Geometry. \/} Springer-Verlag, 1987.


\bibitem[28]{DG} J. Derezinski, C. G\'erard:
{\em Asymptotic completeness in quantum field theory: 
Massive Pauli-Fierz Hamiltonians.} Rev. Math. Phys.
{\bf 11}(4), 383--450 (1999)


\bibitem[29]{DMP} T.C. Dorlas, N. Macris, J.V. Pul\'e:
{\em Localisation in a single-band approximation to random
Schr\"odinger operators in a magnetic field.} Helv. Phys. Acta
{\bf 68}, 329--364 (1995)


\bibitem[30]{EG} P. Elbau, G.M. Graf:
{\em Equality of bulk and edge Hall conductance revisited.}
Commun. Math. Phys.  {\bf 229}, 415--432 (2002)

\bibitem[31]{EGS} A. Elgart, G.M. Graf, J. H. Schenker:
{\em Equality of the bulk and edge Hall conductances
in a mobility gap.} Preprint. {\tt arxiv.org/abs/math-ph/0409017}




\bibitem[32]{El-1} D. Elton: {\sl New examples of zero modes},
J. Phys. A {\bf 33} (41), 7297--7303, (2000)

\bibitem[33]{El-2}  D. Elton: {\sl The local structure of
zero mode producing magnetic potentials}. Commun. Math. Phys. 
{\bf 229}, 121--139 (2002)

\bibitem[34]{E-93} L. Erd\H os: 
{\sl Ground state density of the Pauli operator 
in the large field limit.}  Lett. Math. Phys. {\bf 29}, 219--240 (1993)



\bibitem[35]{E-94} L. Erd\H os:
{\sl Estimates on stochastic oscillatory integrals
on the heat kernel of the magnetic Schr\"odinger operator.}
Duke Math. J. {\bf 76}, 541--566 (1994)




\bibitem[36]{E-95} L. Erd\H os: {\sl Magnetic Lieb-Thirring
inequalities. \/}  Commun. Math. Phys. {\bf 170}, 629--668 (1995)

\bibitem[37]{E-96} L. Erd\H os:
Gaussian decay of the magnetic eigenfunctions,
 Geom. Funct. Anal. (GAFA), {\bf 6} No.2, 231--248 (1996)

\bibitem[38]{E-96(1)} L. Erd\H os:
 {\it Rayleigh-type isoperimetric inequality
with a homogeneous magnetic field.} Calc. Var. and PDE. {\bf 4}, 283--292
(1996)

\bibitem[39]{E-97} L. Erd{\H o}s: {\it Dia- and paramagnetism for
nonhomogeneous magnetic fields.\/} Journal of Math. Phys. {\bf 38}(3),
1289--1317 (1997)


\bibitem[40]{E-98} L. Erd\H os: {\em Lifshitz tail in a magnetic
field: the non-classical regime.} Probab. Theor. Rel. Fields.
{\bf 112}, 321--371 (1998)


\bibitem[41]{E-01} L. Erd\H os: {\em Lifshitz tail in a magnetic
field:  coexistence of classical and quantum behavior
in the borderline case.} Probab. Theor. Rel. Fields.
{\bf 121}, 219--236 (2001)



\bibitem[42]{E-02}  L. Erd{\H o}s: {\it
Spectral shift and multiplicity of the first eigenvalue
of the magnetic Schr\"odinger operator in two dimensions.}
Ann. Inst. Fourier (Grenoble) {\bf 52}, 1833--1874 (2002)



\bibitem[43]{ELV}   L. Erd{\H o}s, M. Loss and V. Vougalter:
{\it Diamagnetic behavior of sums of Dirichlet eigenvalues.}
 Ann. Inst. Fourier (Grenoble), Vol {\bf 50}, no. 3. 891--907 (2000)

\bibitem[44]{ES-I}  L. Erd{\H o}s and J. P. Solovej: {\it Semiclassical
eigenvalue estimates for the Pauli operator with strong
non-homogeneous magnetic fields. I. Non-asymptotic Lieb-Thirring
type estimate.}  Duke Math. J.  {\bf 96}, 127--173 (1999)

\bibitem[45]{ES-II} L. Erd{\H o}s and J. P. Solovej: {\it Semiclassical
eigenvalue estimates for the Pauli operator with strong
non-homogeneous magnetic fields. II. Leading order asymptotic estimates.}
Commun. Math. Phys. {\bf 188}, 599--656 (1997)

\bibitem[46]{ES-III} L. Erd{\H o}s and J. P. Solovej:
{\it The kernel of Dirac operators on $S^3$ and $\bR^3$}.
Rev. Math. Phys. {\bf 13} No. 10, 1247--1280 (2001) 

\bibitem[47]{ES-IV} L. Erd{\H o}s and J. P. Solovej:
{\it Magnetic Lieb-Thirring inequalities with optimal dependence on the field
strength}.  J. Statis. Phys. {\bf 116}, 475--506 (2004)




\bibitem[48]{ES-V} L. Erd{\H o}s and J. P. Solovej:
 {\it Uniform Lieb-Thirring inequality
for the three dimensional Pauli operator with a strong
non-homogeneous magnetic field.}  Ann. Inst. H. Poincar\'e
{\bf 5}, 671--741 (2004)

\bibitem[49]{EV} L. Erd{\H o}s and V. Vougalter:
{\it Pauli operator and Aharonov-Casher theorem for measure 
valued magnetic fields.} Commun. Math. Phys. {\bf 225}, 399--421 (2002)

\bibitem[50]{Fef} C. Fefferman: {\em Stability of Coulomb
systems in a magnetic field.} Proc. Nat. Acad. Sci.USA
{\bf 92}, 5006--5007 (1995)

\bibitem[51]{FLM} W. Fischer, H. Leschke, P. M\"uller:
{\em Spectral localization by Gaussian random potentials
in multi-dimensional continuous space.} J. Stat. Phys. {\bf 101}
935--985 (2000)
 
\bibitem[52]{F1} S. Fournais: {\sl 
Semiclassics of the quantum current in a strong constant magnetic field.}
   Comm. Partial Diff. Eq. {\bf 26} no. 7-8, 1427--1496
(2001)

\bibitem[53]{F2} S. Fournais: {\sl
The magnetisation of large atoms in strong magnetic fields.}
 Commun.  Math. Phys. {\bf 216} no. 2, 375--393 (2001)

\bibitem[54]{F3} S. Fournais: {\sl
Confinement to lowest Landau band and application to quantum current.}
Rev. Math. Phys. {\bf 15}  no. 10, 1219--1253 (2003)


\bibitem[55]{FH} S. Fournais, B. Helffer: {\sl Accurate
estimates for magnetic bottles in connection
with superconducivity.} To appear in Ann. Inst. Fourier.
Preprint in Univ. Texas Mathematical Physics Archive, 04-385 (2004)

\bibitem[56]{Fr} J. Fr\"ohlich: {\em On the infrared problem in a 
model of scalar electrons and massless scalar bosons.}
Ann. Inst. H. Poincar\'e {\bf 19}, 1--103 (1973), and
{\em Existence of dressed one electron states in a class
of persistent models.} Fortschr. Phys. {\bf 22},
159--198 (1974)

\bibitem[57]{FLL} J. Fr\"ohlich, E. H. Lieb and M. Loss, {\em Stability
of Coulomb systems with magnetic fields. I. The one-electron atom. \/} 
Commun. Math. Phys. {\bf 104}, 251--270 (1986)

\bibitem[58]{FGW} J. Fr\"ohlich, G.M. Graf, J. Walcher:
{\em On the extended nature of edge states of quantum Hall Hamiltonians.}
Ann. Henri Poincar\'e {\bf 1}, 405--442 (2000)

\bibitem[59]{FGS1} J. Fr\"ohlich, M. Griesemer, B. Schlein:
{\em Asymptotic completeness for Rayleigh scattering.}
Ann. Henri Poincare {\bf 3}, 107--170 (2002)

\bibitem[60]{FGS2} J. Fr\"ohlich, M. Griesemer, B. Schlein:
{\em Asymptotic completeness for Compton scattering.}
Commun. Math. Phys. {\bf 252}, 415--476 (2004)


\bibitem[61]{FS} J. Fr\"ohlich, T. Spencer: {\em
Absence of diffusion in the Anderson tight binding model for large
disorder or small energy.} Commun. Math. Phys. {\bf 88}
151--184 (1983)


\bibitem[62]{GA} O. Gat, J. Avron: {\em Semiclassical analysis
and the magnetization of the Hofstadter model.}
Phys. Rev. Lett. {\bf 91} 186801 (2003)


\bibitem[63]{GMS} C. G\'erard, A. Martinez, J. Sj\"ostrand:
{\em A mathematical approach to the effective
Hamiltonian in perturbed periodic problems.}
Commun. Math. Phys. {\bf 142}, 217--244 (1991)

\bibitem[64]{GL1} C. G\'erard, I. {\L}aba:
{\em Scattering theory for 3-particle systems
in constant magnetic fields: dispersive case.}
Ann. Inst. Fourier (Grenoble) {\bf 46}, 801--876 (1996)

\bibitem[65]{GL} C. G\'erard, I. {\L}aba:
{\em Multiparticle quantum scattering in constant magnetic fields.}
Math. Surveys and Monographs {\bf 90}, AMS, Providence RI (2002)



\bibitem[66]{GK-01}  F. Germinet, A. Klein:
{\em Bootstrap multiscale analysis and localization in random
media.} Commun. Math. Phys. {\bf 222}, 415--448 (2001)

\bibitem[67]{GK-03}  F. Germinet, A. Klein:
{\em Explicit finite volume criteria for localization
in continuous random media and applications.}
Geom. Funct. Anal. {\bf 13}, 1201--1238 (2003)


\bibitem[68]{GKS} F. Germinet, A. Klein, J. Schenker:
{\em Dynamical delocalization in random Landau Hamiltonians.}
Preprint. {\tt xxx.lanl.gov/math-ph/0412070}


\bibitem[69]{G} F. Ghribi:
{\em Asymptotique de Lifshitz pour des op\'erateurs de
Schr\"odinger magn\'etiques al\'eatoires.}
Ph.D. Thesis. Univ. Paris 13., 2005


\bibitem[70]{GLL} M. Griesemer, E. H. Lieb, M. Loss:
{\sl Ground states in non-relativistic quantum electrodynamics.}
Invent. Math. {\bf 145}, 557--595 (2001)




\bibitem[71]{H-88} B. Helffer:
{\em Effet d'Aharonov Bohm sur un \'etat born\'e de l'\'equation
de Schr\"odinger.} Commun. Math. Phys. {\bf 119},
315--329 (1988)


\bibitem[72]{HM-88} B. Helffer, A. Mohamed: {\sl
Sur le spectre essentiel des op\'erateurs de Schr\"odinger
avec champ magn\'etique.} Ann. Institut Fourier {\bf 38},
95--113 (1988)


\bibitem[73]{HM-96} B. Helffer, A. Mohamed: {\sl
Semiclassical analysis for the ground state
   energy of a Schr\"odinger operator with magnetic wells.}
 J. Funct. Anal. {\bf 138}  no. 1, 40--81 (1996)

\bibitem[74]{HM-04} B. Helffer, A. Morame:
{\sl Magnetic bottles for the Neumann problem: curvature
 effects in the
   case of dimension 3 (general case).} Ann. Sci. \'Ecole Norm. Sup.
 {\bf 37}  no. 1, 105--170 (2004)

\bibitem[75]{HNW} B. Helffer, J. Nourrigat and X. P. Wang:
{\em Sur le spectre de l'\'equation de Dirac (dans ${\bf R}^2$ ou
${\bf R}^3$) avec champ magn\'etique. \/} Ann. scient. \'Ec. Norm.
Sup. $4^{e}$ serie t. 22, 515--533 (1989)




\bibitem[76]{HS-1} B. Helffer, J. Sj\"ostrand: {\sl Analyse
semi-classique pour l'\'equation de Harper I-III.}
 Mem. Soc. Math. France (N.S) {\bf 34} Tome 116 (1989), {\bf 39} Tome 117
(1990), {\bf 40} Tome 118 (1990)

\bibitem[77]{HS-2} B. Helffer, J. Sj\"ostrand: 
{\sl \'Equation de Schr\"odinger avec champ magn\'etique
et \'equation de Harper.} Schr\"odinger operators
(S{\o}nderborg, 1988), 118-197, Lecture Notes in Physics, {\bf 345},
Springer, 1989.

\bibitem[78]{HS-3} B. Helffer, J. Sj\"ostrand: 
{\sl On diamagnetism and de Haas-van Alphen effect.}
Ann. Inst. H. Poincar\'e Phys. Th\'eor. {\bf 52}  no. 4, 303--375
(1990)



\bibitem[79]{HH} R. Hempel, I. Herbst: {\em Strong magnetic
fields, Dirichlet boundaries and spectral gaps.}
Commun. Math. Phys. {\bf 164}, 237--259 (1995)


\bibitem[80]{HSU} H. Hess, R. Schrader and D.A. Uhlenbrock:
{\em Domination of semigroups and generalization of
Kato's inequality.}
Duke Math. J. {\bf 44} no.4, 893-904 (1977)

\bibitem[81]{HK} P. Hislop, F. Klopp:
{\em The integrated density of states for some random operators
with nonsign definite potentials.} J. Funct. Anal. {\bf 195},
12--47 (2002)

\bibitem[82]{HSS} H. Hogreve, R. Schrader and R. Seiler:
{\em A conjecture on spinor functional determinant. \/}
 Nucl. Phys. B {\bf 142}, 525--534 (1978)


\bibitem[83]{HLW} T. Hupfer, H. Leschke, S. Warzel:
{\em Poissonian obstacles with Gaussian walls discriminate
between classical and quantum Lifshits tailing in magnetic fields.}
J. Stat. Phys. {\bf 97}, 725--750 (1999)

\bibitem[84]{Hun} D. Hundertmark: {\em Zur Theorie der magnetischen 
Schr\"odingerhalbgruppe.} Ph.D. Thesis, Bochum, 1997

\bibitem[85]{HS}  D. Hundertmark, B.  Simon:
{\em A diamagnetic inequality for semigroup differences.}
J. reine. ang. Math. {\bf 571}, 107--130 (2004)

\bibitem[86]{HKW} D. Hundertmark, W. Kirsch, S. Warzel:
{\em Classical magnetic Lifshitz tails in three space
dimensions: impurity potential with slow anisotropic decay.}
Markov Proc. Rel. Fields {\bf 9}, 651--660 (2003)


\bibitem[87]{HLMW} T. Hupfer, H. Leschke, P. M\"uller, S. Warzel:
{\em The absolute continuity of the
integrated density of states 
  for magnetic Schr\"odinger operators with certain
unbounded random potentials.} Commun. Math. Phys. {\bf 221},
229--254 (2001)


\bibitem[88]{IK} T. Ikebe, T. Kato: {\em
Uniqueness of the self-adjoint extension of singular
elliptic differential operators.} Arch. Rat. Mech. Anal.
{\bf 9}, 77-92 (1962)

\bibitem[89]{I} V. Ivrii: {\rm Microlocal Analysis and Precise
Spectral Asymptotics.} Springer, 1998.

\bibitem[90]{I-05} V. Ivrii: {\em Sharp spectral asymptotics
for operators with irregular coefficients. III, IV. V.} Preprints 2005.
{\tt xxx.lanl.gov/math.AP/0510326, 05103267, 0510328}


\bibitem[91]{Iw} H. Iwashita: {\em On the long-range scattering
for one- and two-particle Schr\"odinger operators
with constant magnetic fields.} Tsukuba J. Math. {\bf 19}, 369--376 
(1995)

\bibitem[92]{Iwa} A. Iwatsuka: {\em Examples of absolutely continuous
Schr\"odinger operators in magnetic fields.}
Publ. RIMS {\bf 21}, 385--401 (1985)

\bibitem[93]{K-73} T. Kato:
{\em Schr\"odinger operators with singular potentials. \/}
Israel J. Math. {\bf 13}, 135-148 (1973)


\bibitem[94]{K} T. Kato: {\em Remarks on Schr\"odinger operators
with vector potentials. \/} Integral Eq. Operator Theory {\bf 1},
 103--113 (1978)

\bibitem[95]{KNNN} F. Klopp, S. Nakamura, F. Nakano, Y. Nomura:
{\em Anderson localization for 2D discrete Schr\"odinger operators
with random magnetic fields.}
Ann. Henri Poincar\'e {\bf 4} 795--811 (2003)


\bibitem[96]{KR}  F. Klopp, G. Raikov: 
{\em Lifshitz tails in constant magnetic fields.} Preprint 2005,
{\tt xxx.lanl.gov/math-ph/0509022}

\bibitem[97]{La-93} I. {\L}aba: {\em Long-range one-particle
scattering in a homogeneous magnetic field.}
Duke Math. J. {\bf 70}, 283--303 (1993) 

\bibitem[98]{La-95} I. {\L}aba: {\em Scattering for hydrogen-like
systems in a constant magnetic field.}
Comm. in P.D.E {\bf 20}, 741--762 (1995)


\bibitem[99]{Last} Y. Last: {\sl Zero measure spectrum for
the almost Mathieu operator.} Commun. Math. Phys. {\bf 164}
421--432 (1994)

\bibitem[100]{LaS} Y. Last, B. Simon: {\em The essential spectrum
of Schr\"odinger, Jacobi, and CMV operators.}
Preprint. Univ. Texas Mathematical Physics Archive
05-112. (2005)

\bibitem[101]{Lein} H. Leinfelder: {\em Gauge invariance 
of Schr\"odinger operators and related spectral properties},
J. Op. Theory, {\bf 9}, 163--179 (1983)

\bibitem[102]{LS-81} H. Leinfelder, C. Simader:
{\em Schr\"odinger operators with singular magnetic vector potentials. }
Math Z., {\bf 176}, 1--19 (1981)

\bibitem[103]{LMW} H. Leschke, P. M\"uller, S. Warzel:
{\em A survey of rigorous results on random Schr\"odinger operators
for amorphous solids.} Markov. Proc. Rel. Fields {\bf 9}, 729--760 (2003)






\bibitem[104]{L-92} E. H. Lieb: {\em The flux
phase problem on planar lattices.}
Helv. Phys. Acta {\bf 65}, 247--255 (1992)


\bibitem[105]{L-94} E. H. Lieb:  {\em  The flux phase
 of the half-filled band. \/} Phys. Rev. Lett.  {\bf 73},
2158--2161 (1994)

\bibitem[106]{L-04} E. H. Lieb: {\em Quantum mechanics, the stability
of matter and quantum electrodynamics.} Jahresber. Deutsch. Math.-Verein.
{\bf 106}, 93--110 (2004)


\bibitem[107]{LL-86} E. H. Lieb, M. Loss: {\em
Stability of Coulomb systems with magnetic fields  II.}
Commun. Math. Phys, {\bf 104} 271--282 (1986)

\bibitem[108]{LL-93} E. H. Lieb, M. Loss: {\em
Fluxes, Laplacians, and Kasteleyn's theorem.}
 Duke Math. J. {\bf 71}
  no. 2, 337--363 (1993)

\bibitem[109]{LL-02} E. H. Lieb, M. Loss: {\sl
Stability of a model of relativistic quantum electrodynamics.}
Commun. Math. Phys. {\bf 228}, 561--588 (2002)


\bibitem[110]{LL-03} E. H. Lieb, M. Loss: {\sl
Existence of atoms and molecules in non-relativistic
quantum electrodynamics.}
Adv. Theor. Math. Phys. {\bf 7}, 667-710 (2003)


\bibitem[111]{LL-05} E. H. Lieb, M. Loss: {\sl
The thermodynamic limit for matter interacting with Coulomb forces
and with the quantized electromagnetic field: I. The lower bound.}
Commun. Math. Phys. {\bf 258}, 675--695 (2005)

\bibitem[112]{LLSi} E. H. Lieb, M. Loss and H. Siedentop: {\sl 
Stability of relativistic matter via Thomas-Fermi theory.}
Helv. Phys. Acta {\bf 69}, 974--984 (1996)


\bibitem[113]{LLS} E. H. Lieb, M. Loss and J. P. Solovej: {\sl 
Stability of Matter in Magnetic Fields}, Phys. Rev. Lett. {\bf 75},
 985--989 (1995)


\bibitem[114]{LSS} E. H. Lieb, H. Siedentop and J. P. Solovej:
{\em Stability and instability of realativistic electrons in classical 
electromagnetic fields.} J. Stat. Phys. {\bf 89} 37--59 (1997)



\bibitem[115]{LS} E. H. Lieb, B. Simon: {\em The Thomas-Fermi
theory of atom, molecules and solids.} Adv. in Math. {\bf 23} 22--116, (1977)


\bibitem[116]{LSY-I} E. H. Lieb, J. P. Solovej and J. Yngvason:
{\sl Asymptotics of heavy atoms in high magnetic fields: I. Lowest
Landau band region}, Commun. Pure  Appl. Math. {\bf 47},
513--591 (1994)

\bibitem[117]{LSY-II} E. H. Lieb, J. P. Solovej and J. Yngvason:
{\sl Asymptotics of heavy atoms in high magnetic fields: II. Semiclassical
regions. \/}  Commun. Math. Phys. {\bf 161}, 77--124 (1994)

\bibitem[118]{LSY-III} E. H. Lieb, J. P. Solovej and J. Yngvason:
{\em Ground states of large quantum dots in magnetic fields. \/}
Phys. Rev. B {\bf 51}, 10646--10665 (1995)




\bibitem[119]{LT}  E. H. Lieb, W. Thirring: {\sl Inequalities for moments
of the eigenvalues of the Schr\"odinger Hamiltonian and their relation
to Sobolev inequalities.} In: Studies in Mathematical Physics
(E. Lieb, B. Simon, A. Wightman eds.) Princeton University Press,
269--330 (1975)



\bibitem[120]{LYa} E. H. Lieb, H.T. Yau:
{\sl The stability and instability of relativistic matter.}
Commum. Math. Phys. {\bf 118}, 177--213 (1988)



\bibitem[121]{LT-87} M. Loss and B. Thaller:
{\sl Scattering of particles by long-range 
magnetic fields.} Ann. Physics {\bf 176}, 159--180 (1987)

\bibitem[122]{LT-88} M. Loss and B. Thaller:
{\sl Short range scattering in long-range  magnetic fields: the
relativistic case.} J. Diff. Equ. {\bf 73}, 225--236 (1988)


\bibitem[123]{LT-97} M. Loss and B. Thaller: {\sl Optimal
heat kernel estimates for Schr\"odinger operators
with magnetic fields in two dimensions.} Commun. Math. Phys. 
{\bf 86}  no. 1, 95--107 (1997)


\bibitem[124]{LY} M. Loss and H.-T. Yau: {\sl Stability of Coulomb
systems with magnetic fields: III. Zero energy bound states of
the Pauli operator. \/} Commun. Math. Phys. {\bf 104}, 283--290 (1986)

\bibitem[125]{Ma} N. Macris: {\em Spectral flow and level spacing
of edge states for quantum Hall Hamiltonians.} J. Phys. A: Math. Gen.
{\bf 36}, 1565--1581 (2003)


\bibitem[126]{Ma-86} P. Malliavin: {\em Minoration de l'etat
fondamental de l'\'equation de Schr\"odinger du magn\'etisme et calcul
 des variations} C. R. Acad. Sci.
{\bf 302}, 481--486 (1986)

\bibitem[127]{MU} H. Matsumoto, N. Ueki: {\em Spectral analysis of 
Schr\"odinger operators with
   magnetic fields.} J. Funct. Anal. {\bf 140}, no. 1, 218--255
(1996)




\bibitem[128]{Mi} K. Miller: {\em Bound States of Quantum Mechanical
Particles in Magnetic Fields. \/} Ph.D. Thesis, Princeton University, 1982.

\bibitem[129]{MS} K. Miller, B. Simon:
{\em Quantum magnetic Hamiltonians with remarkable spectral
properties.} Phys. Rev. Lett. {\bf 44}, no. 25, 1706-1707 (1980)


\bibitem[130]{Mon} R. Montgomery: {\em Hearing the zero locus
of a magnetic field.} Commun. Math. Phys. {\bf 168} 651--675 (1995)

\bibitem[131]{Mo} A. Morame: {\em Absence of singular
spectrum for a perturbation of a two-dimensional Laplace-Beltrami
operator with periodic electromagnetic potential.}
J. Phys. A {\bf 31}, no. 37,  7593-7601 (1998)



\bibitem[132]{Na1} S. Nakamura:
{\em Gaussian decay estimates for the eigenfunctions of 
magnetic Schr\"odinger
   operators.}
 Comm. Partial Diff. Eq {\bf 21}  no. 5-6, 993--1006 (1996)

\bibitem[133]{Na2} S. Nakamura:
{\em Lifshitz tail for 2D discrete Schr\"odinger operator
with random magnetic field.}
Ann. Henri Poincar\'e {\bf 1}, 823--835 (2000)

\bibitem[134]{Na3} S. Nakamura:
{\em Lifshitz tail for  Schr\"odinger operator
with random magnetic field.}
Commun. Math. Phys. {\bf 214}, 565--572 (2000)

\bibitem[135]{N} G. Nenciu: {\em  Dynamics of band electrons in
electric and magnetic fields: rigorous justificatino of the
effective Hamiltonians.} Rev. Mod. Phys. {\bf 63}, 91--127 (1991)

\bibitem[136]{PST} G. Panati, H. Spohn, S. Teufel: {\em Effective
dynamics for Bloch electrons: Peierls
substitution and beyond.}  Commun. Math. Phys. {\bf 242},  547--578 (2003)


\bibitem[137]{Pu} J. Puig: {\em Cantor
spectrum for the almost Mathieu operator. } Commun. Math. Phys.
{\bf 244}, 297--309 (2004)

\bibitem[138]{RW} G. Raikov, S. Warzel: {\sl Quasi-classical versus
non-classical spectral asymptotics for magnetic Schr\"odinger operators
with decreasing electric potentials.} Rev. Math. Phys. {\bf 14},
1051--1072 (2002), and
{\sl 
Spectral asymptotics for magnetic Schr\"odinger
   operators with rapidly decreasing electric potentials.}
 C. R. Math. Acad. Sci. Paris {\bf 335} no. 8, 683--688 (2002)

\bibitem[139]{RSi} M. Reed, B. Simon: 
{\em Methods of Modern Mathematical Physics. I--IV.}
Academic Press, 1975, 1978, 1980



\bibitem[140]{R} D. Robert: {\sl On scattering theory for long range
perturbations of Laplace operators.} J. Anal. Math. {\bf 59}, 189--203 (1992)

\bibitem[141]{RS} G. Rozenblum, N. Shirokov: 
{\em Infiniteness of zero modes for the Pauli operator
with singular magnetic field.} Preprint 2005,
 {\tt xxx.lanl.gov/math-ph/0501059}


\bibitem[142]{SKR} H. Schulz-Baldes, J. Kellendonk, T. Richter:
{\em Simultaneous quantization of edge and bulk Hall conductivity.}
J. Phys. A: Math. Gen. {\bf 33} L27--L33 (2000)

\bibitem[143]{S} R. Seiringer: {\sl On the maximal ionization
of atoms in strong magnetic fields.} J. Phys. A {\bf 34}, 1943--1948
(2001)


\bibitem[144]{Sh-96} Z. Shen: {\sl Eigenvalue asymptotics
and exponential decay of eigenfunctions for Schr\"odinger operators
with magnetic fields.} Trans. Amer. Math. Soc. {\bf 348}, 4465--4488
(1996)

\bibitem[145]{Sh-99} Z. Shen: {\sl On the moments of negative eigenvalues
for the Pauli operator}, J. Diff. Eq. {\bf 149}, 292--327 (1998)
and {\bf 151}, 420--455 (1999)




\bibitem[146]{S-77} B. Simon: {\em An abstract Kato's inequality
for generators of positivity preserving semigroups.\/}
Indiana Univ. Math. J. {\bf 26}, 1067--1073 (1977)

\bibitem[147]{S-79} B. Simon: {\rm Functional Integration and Quantum
Physics. \/} Second Edition, AMS Chelsea Publishing, 2005


\bibitem[148]{S-79(a)} B. Simon: {\em Maximal and minimal Schr\"odinger
forms. \/} J. Operator Theory {\bf 1}, 37--47 (1979)

\bibitem[149]{S-79(b)} B. Simon:
{\em Kato's inequality and the comparison semigroups.}
J. Funct. Anal. {\bf 32}, 97--101 (1979)

\bibitem[150]{S-79(c)} B. Simon: 
{\em Phase space analysis of simple scattering systems: extension
of some work of Enss.} Duke Math. J. {\bf 46}, 119--168 (1979)




\bibitem[151]{S-82} B. Simon: {\em Schr\"odinger semigroups. \/}
Bull. Am. Math. Soc. {\bf 7}, 447--526 (1982)

\bibitem[152]{S-00} B. Simon: {\em Schr\"odinger operators in the
twentieth century. \/} J. Math. Phys. {\bf 41}, 3523--3555 (2000)






\bibitem[153]{Sk} E. Skibsted: {\em Asymptotic completeness for particles
in combined constant electric and magnetic fields. II.}
Duke Math. J. {\bf 89}, 307--350 (1997)




\bibitem[154]{So-94} A. Sobolev: {\sl The quasi-classical
asymptotics of local Riesz means for the Schr\"odinger operator
in a strong homogeneous magnetic field.} Duke Math. J.
{\bf 74}, 319--429 (1994)




\bibitem[155]{So-96}  A. Sobolev: {\sl
Two-term asymptotics for the sum of eigenvalues of the
   Schr\"odinger operator with Coulomb singularities
 in a homogeneous magnetic field.} Asymptotic Anal. {\bf 13} no. 4,
   393--421 (1996)

\bibitem[156]{So-97}
 A. Sobolev: {\sl  Lieb-Thirring inequalities for the Pauli operator
in three dimensions}, IMA Vol. Math. Appl. {\bf 95}, 155--188  (1997)

\bibitem[157]{So-98} A. Sobolev: {\sl Quasiclassical asymptotics for the
Pauli operator}, Commun. Math. Phys. {\bf 194},  109--134 (1998)




\bibitem[158]{So-99} A.V. Sobolev: {\em
Absolute continuity of the periodic magnetic Schr\"odinger
   operator.} Invent. Math. {\bf 137} no. 1, 85--112 (1999)




\bibitem[159]{Sor} V. Sordoni:
{\em Gaussian decay for the eigenfunctions of a Schr\"odinger operator with
   magnetic field constant at infinity.}
 Comm. Partial Diff. Eq. {\bf 23} no. 1-2, 223--242 (1998)


\bibitem[160]{Tam} H. Tamura: {\em 
Asymptotic distribution of eigenvalues for Schr\"odinger operators with
   magnetic fields.} Nagoya Math. J. {\bf 105}, 49--69 (1987)


\bibitem[161]{Thom}  L. E. Thomas: {\em Time dependent approach
to scattering from impurities in a crystal.} Commun. Math. Phys.
{\bf 33}, 335--343 (1973)

\bibitem[162]{U-94} N. Ueki: {\em Lower bounds for the spectra
of Schr\"odinger operators with magnetic fields.}
J. Funct. Anal. {\bf 120}, 344--379 (1994), Erratum: J. Funct.
Anal. {\bf 127}, 257--258 (1995)



\bibitem[163]{U-02} N. Ueki: {\em Estimates on the heat
kernel of the Pauli Hamiltonian and its application
to problems on hypoellipticity of the
$\overline{\partial_b}$-Laplacian.} Math. Z. {\bf 239}(1),
69--97 (2002)



\bibitem[164]{W-97} W.-M. Wang: {\em Microlocalization, percolation
and Anderson localization for the magnetic Schr\"odinger operator
with a random potential.} J. Funct. Anal. {\bf 146} 1--26 (1997)

\bibitem[165]{W-00} W.-M. Wang: {\em Supersymmetry and density of states
of the magnetic Schr\"odinger operator with a random potential
revisited.} Comm. Part. Diff. Eq. {\bf 25} 601--679 (2000)

\bibitem[166]{Wil} M. Wilkinson: {\sl An exact renormalisation group
for Bloch electrons in a magnetic field.} J. Phys. A {\bf 20} 1791 (1987)



\end{thebibliography}
\end{document}